\documentclass[onecolumn,showkeys,preprintnumbers,amsmath,amssymb]{revtex4-2}

\usepackage{graphicx}
\usepackage{dcolumn}
\usepackage{bm}
\usepackage[utf8]{inputenc}
\usepackage{amsmath}
\usepackage{amssymb}
\usepackage{booktabs}
\usepackage{color}
\usepackage{units}	
\usepackage{subcaption}
\usepackage{float}
\usepackage{svg}
\usepackage{hyperref}
\usepackage{changepage}
\usepackage{comment}
\usepackage{microtype}
\begin{document}
\author{Marcelo H. Alvarenga}
 \altaffiliation{}
 \email{marcelo.alvarenga@ufla.br}
\affiliation{Departamento de Física, UFLA, Lavras, MG, Brazil
}%
\author{Júlio C. Fabris}
 \altaffiliation{}
 \email{julio.fabris@cosmo-ufes.org}
\affiliation{Núcleo Cosmo-ufes \& Departamento de Física, UFES, Vitória, ES, Brazil
\\
National Research Nuclear University MEPhI, Kashirskoe sh. 31, Moscow 115409, Russia
}%
\author{Rodrigo Santos Bufalo}
\altaffiliation{}
\email{rodrigo.bufalo@ufla.br}
\affiliation{Departamento de Física, UFLA, Lavras, MG, Brazil
}%

\date{\today}

\begin{abstract}
The absence of an algebraic relation between the Ricci scalar and the trace of the energy-momentum tensor in Non-Conservative Unimodular Gravity ($\mathrm{NUG}$) opens new possibilities for constructing spacetime geometries from prescribed curvature profiles. Exploiting this property, we develop a curvature reconstruction framework in which traversable wormhole geometries are generated directly from a prescribed curvature profile rather than from the matter sector or the metric functions. The reconstruction procedure is implemented for a power-law Ricci scalar combined with three different redshift functions, leading to distinct classes of static and spherically symmetric wormhole solutions. We investigate their geometrical properties, asymptotic behavior and null energy condition. Although the radial null energy condition remains necessarily violated at the throat, we show that the dimensionless curvature parameter controls the intensity of the exotic matter supporting the wormhole, while the curvature exponent determines the global asymptotic structure of the reconstructed spacetime. In particular, the critical configuration separating asymptotically flat solutions from geometries with residual asymptotic deformation naturally emerges from the reconstruction scheme. These results establish curvature reconstruction from prescribed Ricci scalar profiles as a viable geometrical framework for generating compact-object spacetimes within non-conservative unimodular gravity.
\end{abstract}

\title{Wormhole Reconstruction in Non-Conservative Unimodular Gravity: Ricci Scalar as an Independently Prescribed Curvature Profile.}
\maketitle
\section{Introduction.}
Traversable wormholes constitute one of the most remarkable structures predicted by General Relativity ($\mathrm{GR}$), representing hypothetical geometrical bridges connecting distinct regions of spacetime \cite{Morris:1988cz, Visser:1995cc}. Their conceptual origins can be traced back to the Einstein-Rosen bridge \cite{PhysRev.48.73}, although the modern notion of traversability was established much later by Morris and Thorne \cite{Morris:1988cz}. Beyond their mathematical elegance, wormholes have attracted sustained attention because they provide a natural framework for investigating fundamental aspects of gravitation, spacetime topology, causality, and the interplay between geometry and matter.

Historically, the modern theory of traversable wormholes emerged from an unusual interaction between science and science communication. While preparing his novel \emph{Contact}, Carl Sagan consulted Kip Thorne about the possibility of interstellar travel consistent with $\mathrm{GR}$. This discussion ultimately motivated Michael Morris and Kip Thorne to formulate the first traversable wormhole solution satisfying well-defined geometrical conditions \cite{Thorne1994, Sagan1985, Morris:1988cz}. Since then, the Morris-Thorne geometry has become the canonical framework for the study of traversable wormholes.

Within the Morris-Thorne framework, a static and spherically symmetric traversable wormhole is completely characterized by two metric functions: the redshift function $\Phi(r)$ and the shape function $b(r)$. The throat corresponds to a hypersurface of minimum areal radius and must satisfy the flare-out condition
\begin{align}
b'(r_0)<1,
\end{align}
where $r_0$ is throat radius, which guarantees the traversability of the geometry. Up to this stage, these conditions are purely geometrical and independent of the underlying gravitational dynamics.

The situation changes once these geometrical requirements are combined with Einstein's field equations. In $\mathrm{GR}$, the flare-out condition necessarily implies the violation of the Null Energy Condition (NEC) at, or in the vicinity of, the throat \cite{Morris:1988cz}. Consequently, traversable wormholes require the presence of exotic matter, i.e., matter sources satisfying $\rho+p_r<0$. This result has motivated extensive research over the past decades, including models supported by phantom scalar fields, effective fluids, quantum corrections and, more recently, modified theories of gravity, where the exotic character of the matter content can be partially or entirely transferred from to the gravitational sector \cite{Ellis:1973yv, Bronnikov:1973fh, Visser:1995cc, Lobo:2005us,  PhysRevD.87.067504}. Another fundamental question concerns the dynamical stability of traversable wormholes. Besides satisfying the geometrical and energy requirements, physically viable configurations should remain stable under small perturbations. Accordingly, stability analyses based on thin-shell formalisms, linear perturbation theory and modified gravitational scenarios have become an active topic in the wormhole literature \cite{Visser:1995cc, Poisson:1995sv, Gonzalez:2008wd, PhysRevD.71.124022}.

Among the various modified theories of gravity, Unimodular Gravity ($\mathrm{UG}$) has attracted considerable attention owing to its distinct variational formulation and its long-standing connection with the cosmological constant problem \cite{RevModPhys.61.1, Anderson:1971pn, 1991JMP....32.1337N, Ellis:2010uc}. In $\mathrm{UG}$, the determinant of the metric is fixed at the variational level, leading to trace-free gravitational field equations. This formulation has been extensively investigated in cosmological scenarios \cite{Godel:2025, Alvarez:2023eqo, Ellis:2013uxa} and has also been employed in the construction of traversable wormhole solutions \cite{Cataldo_2026, Pastén_2026}. The trace-free equations do not determine the Ricci scalar independently, so that the Ricci scalar $R$ is not algebraically related to the trace of the energy-momentum tensor $T$. Moreover, unlike in $\mathrm{GR}$, the contracted Bianchi identities do not automatically imply the usual covariant conservation of the energy-momentum tensor $T_{\mu\nu}$.

In most applications of conservative $\mathrm{UG}$, the usual conservation law,
\begin{align}
\nabla_{\mu}T^{\mu\nu}=0,
\end{align}
is imposed as an additional condition to close the system of field equations. This immediately restores Einstein's equations with a cosmological constant emerging as an integration constant, simultaneously recovering the algebraic relation between $R$ and $T$. Consequently, the scalar sector becomes algebraically closed in the conservative formulation. 

Motivated by this observation, a non-conservative extension of unimodular gravity has recently been proposed, where the usual conservation law is no longer imposed as an external constraint. Since its introduction, this framework has been successfully applied to several cosmological scenarios, spherically symmetric configurations and compact objects, revealing novel phenomenological consequences absent in the conservative formulation \cite{universe9120515, aguilarpérez2026, Alencar:2026ffr, Alvarenga:2025nwe}. More recently, $\mathrm{NUG}$ has also been employed in the construction of exact wormhole solutions, where the non-conservative sector gives rise to a dynamical cosmological function coupled to scalar and Maxwell fields \cite{alencar2026}. 

Beyond these applications, the non-conservative formulation possesses another remarkable consequence that has remained essentially unexplored. Since the Ricci scalar is no longer algebraically constrained by the matter trace, it can instead be prescribed independently. This property fundamentally distinguishes the non-conservative theory from both $\mathrm{GR}$ and conservative $\mathrm{UG}$, where the scalar curvature is ultimately determined by the matter content. Therefore, the present work explores a complementary consequence of treating the Ricci scalar as an independent geometric quantity. Rather than prescribing the matter content or the metric functions, we take the Ricci scalar as the primary geometric input and reconstruct the wormhole geometry directly from a prescribed curvature profile.

It is important to emphasize that the reconstruction procedure developed here relies exclusively on the trace-free gravitational field equations together with the geometrical definition of the Ricci scalar. The non-conservative formulation does not modify the reconstruction equations themselves; instead, it provides the theoretical consistency required to treat the Ricci scalar as an independently prescribed geometric quantity. Unlike previous approaches, which typically prescribe the matter sector, metric functions, or an effective cosmological function, the present work reconstructs the wormhole geometry directly from a prescribed curvature profile.

In this work, we investigate three representative wormhole geometries reconstructed from different choices of the redshift function under a common power-law curvature profile, $R(r)=R_0\left(\frac{r_0}{r}\right)^n$.
For each configuration, the pair $\{R(r),\Phi(r)\}$ completely determines the wormhole geometry through the reconstruction of the shape function $b(r)$, while the trace-free gravitational field equations determine the null-energy combinations $\left(\rho+p_r\right)$ and $\left(\rho+p_t\right)$. We analyze the reconstructed solutions from both geometrical and physical perspectives, with particular emphasis on the NEC, their asymptotic behavior, and the influence of the curvature exponent $n$ together with the dimensionless curvature parameter $\lambda\equiv R_0r_0^2$.

As pointed out in \cite{Cataldo_2026}, the trace-free field equations alone do not eliminate the necessity of exotic matter for supporting traversable wormholes. Our results corroborate this conclusion. Nevertheless, we demonstrate that the dimensionless curvature parameter $\lambda$ plays a central role in controlling the magnitude of NEC violation in the reconstructed geometries. As $\lambda$ approaches its critical value ($\lambda\simeq2$), the violation of the NEC near the throat becomes progressively weaker, indicating that the curvature sector significantly reduces the exotic matter content supporting the wormhole.

The remainder of this paper is organized as follows. In Sec.~\ref{sec:2} presents the theoretical framework of non-conservative unimodular gravity. In Sec.~\ref{sec:3}, we derive the field equations for static and spherically symmetric wormholes and formulate the curvature reconstruction method. In Sec.~\ref{sec:4} is devoted to the construction of wormhole geometries obtained from a power-law curvature profile. Their geometrical and physical properties are investigated through a detailed numerical analysis in Sec.~\ref{sec:5}. Finally, Sec.~\ref{sec:6} summarizes our main results and discusses future perspectives.

\section{Non-conservative unimodular gravity.}\label{sec:2}

The field equations in Non-conservative Unimodular Gravity ($\mathrm{NUG}$)
are given by
\begin{align}
R_{\mu\nu}-\frac{1}{4}g_{\mu\nu}R & =8\pi G\left(T_{\mu\nu}-\frac{1}{4}g_{\mu\nu}T\right)\label{eq:1}\\
\frac{R_{;\nu}}{4} & =8\pi G\left(T_{\mu\nu}^{\,\,;\mu}-\frac{T_{;\nu}}{4}\right).\label{eq:2}
\end{align}
The Eq. (\ref{eq:2}) can be rewritten according to
\begin{align}
T_{\mu\nu}^{\,\,;\mu} & =\frac{R_{;\nu}}{32\pi G}+\frac{T_{;\nu}}{4}.\label{eq:3}
\end{align}
The Eq.~(\ref{eq:3}) is the result of applying Bianchi's
identities to (\ref{eq:1}). It represents an equation where the usual
conservation of the energy-momentum tensor is not satisfied. The Eqs.~
(\ref{eq:1}) and (\ref{eq:3}) can be mapped onto $\mathrm{GR}$
with a dynamic cosmological constant, in this sense, by choosing $\Lambda\left(R,T\right)\equiv\frac{1}{4}\left(R+8\pi GT\right)$,
we then obtain $\mathrm{GR+\Lambda\left(r\right)}$ (the dependence
is only on r due to static and spherically symmetric symmetry). It
is worth noting that in this work, we did not choose such a definition;
we will address the equations in $\mathrm{NUG}$ as they appear in
(\ref{eq:1}) and (\ref{eq:3}).
\section{Wormhole Equations and Curvature Reconstruction.}\label{sec:3}
\subsection{Field Equations for Static and Spherically Symmetric Wormholes.}
The spherically symmetric and static metric that describes the geometry
of the wormhole is given by
\begin{align}
ds^{2} & =-e^{2\Phi\left(r\right)}dt^{2}+\frac{dr^{2}}{1-\frac{b\left(r\right)}{r}}+r^{2}d\Omega^{2},\label{eq:4}
\end{align}
where $r$ is the areal radial coordinate and $d\Omega^{2}=d\theta^{2}+\sin^{2}\theta\,d\varphi^{2}$. Each asymptotic region is covered by $r\in[r_{0},\infty)$, with $r=r_{0}$ denoting the throat. The functions $b(r)$ and $\Phi(r)$ are the shape and redshift functions, respectively. 

For the geometry to represent a traversable wormhole, the shape function
must satisfy a set of geometrical conditions,
\begin{align}
b(r_{0})&=r_{0}, \\
b'(r_{0})&<1, \\
b'(r)&<\frac{b(r)}{r}, \\
b(r)&<r,
\end{align}
throughout the domain $r\geq r_{0}$ \cite{PhysRevD.56.4745, Morris:1988cz}.
The first condition identifies $r_0$ as the minimum of the areal radius,
and therefore locates the wormhole throat, i.e., the spherical
two-surface of minimum area connecting the two asymptotic regions \cite{Morris:1988cz}. The inequality $b(r)<r$ guarantees
that the radial metric coefficient remains positive away from the throat,
allowing the two sides of the wormhole to extend from the minimum-radius
surface toward their respective asymptotic regions. Most importantly, the
condition
\begin{align}
b'(r_{0})<1
\end{align}
is the flare-out condition. It ensures that the areal radius increases away from the throat on both sides, so that the throat represents a local minimum of the areal radius and the spatial geometry opens outward. This condition is therefore the fundamental geometrical requirement for a traversable wormhole.

Using the metric (\ref{eq:4}) in the Eqs.~(\ref{eq:1}) and
(\ref{eq:2}) we obtain
\begin{align}
2\left(1-\frac{b}{r}\right)\Phi'{}^{2}+\frac{4}{r}\left(1-\frac{3}{4}\frac{b}{r}\right)\Phi'+2\left(1-\frac{b}{r}\right)\Phi''-\frac{1}{r}\left(\Phi'b'-\frac{2}{r}b'\right)=\kappa\left[p_{r}+2p_{t}+3\rho\right]\label{eq:5}\\
\nonumber \\-2\left(1-\frac{b}{r}\right)\Phi'{}^{2}+\frac{4}{r}\left(1-\frac{5}{4}\frac{b}{r}\right)\Phi'-2\left(1-\frac{b}{r}\right)\Phi''+\frac{1}{r}\left(\Phi'b'+\frac{2}{r}b'\right)-\frac{4b}{r^{3}}=\kappa\left[3p_{r}-2p_{t}+\rho\right]\label{eq:6}\\
\nonumber \\\frac{1}{r}\left(b'-\frac{b}{r}\right)\Phi'-2\left(1-\frac{b}{r}\right)\Phi'{}^{2}-2\left(1-\frac{b}{r}\right)\Phi''-\frac{2b}{r^{3}}=\kappa\left[p_{r}-2p_{t}-\rho\right]\label{eq:7}\\
\nonumber \\\frac{R'}{4\kappa}=\left(p_{r}+\rho\right)\Phi'+\frac{2}{r}\left(p_{r}-p_{t}\right)+\frac{3}{4}p_{r}^{'}-\frac{1}{2}p_{t}^{'}+\frac{\rho^{'}}{4},\label{eq:8}
\end{align}
where the line represents the derivative with respect to the radial
coordinate $r$ and $\kappa=8\pi G$. We used as a source an anisotropic
fluid whose energy-momentum tensor is given by
\begin{align}
T_{\mu\nu} & =\left(\rho+p_{t}\right)u_{\mu}u_{\nu}+p_{t}g_{\mu\nu}+\left(p_{r}-p_{t}\right)v_{\mu}v_{\nu},\label{eq:9}
\end{align}
where $u_{\mu}$ is the $4$-velocity rates that satisfy $u_{\mu}u^{\mu}=-1$;
and $v_{\mu}$ is a space-type auxiliary vector that satisfy $v_{\mu}v^{\mu}=1$.
In particulary, by choosing comoving observers, we have
\begin{align}
u^{\mu} & =\left(e^{-\Phi\left(r\right)},\,0,\,0,\,0\right),\label{eq:10}\\
v^{\mu} & =\left(0,\,\sqrt{1-\frac{b}{r}},\,0,\,0\right),\label{eq:11}
\end{align}
were we use the metric (\ref{eq:4}) and note that $u^{\mu}$, $v^{\mu}$
are orthogonal: $u^{\mu}v_{\mu}=0$.

The Eqs.~(\ref{eq:5})-(\ref{eq:8}) form an underdetermined
system, an intrinsic characteristic of unimodular theory. Of the three
trace-free field equations (\ref{eq:5})-(\ref{eq:7}), only two are
independent. Although Eq.~(\ref{eq:8}) originates from the non-conservation law, it is not independent from the trace-free field equations and reduces identically to $0=0$ once Eqs.~(\ref{eq:5})-(\ref{eq:7}) are used. Consequently, the theory does not provide a fundamental relation between the Ricci scalar and the matter trace, $R$ and $T$, leaving one arbitrary function unspecified. Therefore, an additional condition is required to close the system.

We can rewrite the Eqs.~(\ref{eq:5})-(\ref{eq:7}) in a compact
form, as follows
\begin{align}
\kappa\rho & =\frac{1}{3}\left[\frac{4b'}{r^{2}}-R-\kappa p_{r}-2\kappa p_{t}\right]\label{eq:12}\\
\kappa p_{r} & =\frac{1}{3}\left[R+\frac{8}{r}\left(1-\frac{b}{r}\right)\Phi'-\frac{4b}{r^{3}}+2\kappa p_{t}-\kappa\rho\right]\label{eq:13}\\
\kappa p_{t} & =\frac{1}{2}\left[\kappa p_{r}-\kappa\rho-R-\frac{4}{r}\left(1-\frac{b}{r}\right)\Phi'+\frac{2b'}{r^{2}}+\frac{2b}{r^{3}}\right],\label{eq:14}
\end{align}
where the geometric definition for the curvature scalar $R$ was used
in (\ref{eq:5})-(\ref{eq:7}), given by
\begin{align}
R & =-2\left(1-\frac{b}{r}\right)\Phi'{}^{2}-\frac{4}{r}\left(1-\frac{3}{4}\frac{b}{r}\right)\Phi'-2\left(1-\frac{b}{r}\right)\Phi''+\frac{1}{r}\left(\Phi'b'+\frac{2}{r}b'\right).\label{eq:16-1}
\end{align}
We can combine the Eqs.~(\ref{eq:13}) and (\ref{eq:14}) to
obtain the following expressions
\begin{align}
\kappa p_{r} & =\frac{2}{r}\left(1-\frac{b}{r}\right)\Phi'-\frac{b}{r^{3}}+\frac{b'}{r^{2}}-\kappa\rho\label{eq:15}\\
\kappa p_{t} & =\frac{b}{2r^{3}}+\frac{3b'}{2r^{2}}-\frac{R}{2}-\frac{1}{r}\left(1-\frac{b}{r}\right)\Phi'-\kappa\rho.\label{eq:16}
\end{align}
The Eqs.~(\ref{eq:15}) and (\ref{eq:16}) were analyzed in
the paper \cite{Cataldo_2026}. A direct analysis of these expressions
shows that the Null Energy Condition (NEC) in the context of Unimodular
Gravity ($\mathrm{UG}$) continues to be violated in the throat. We
can observe this property in the Eq.~(\ref{eq:15}), which,
when evaluated in the throat, gives us
\begin{align}
\left(\rho+p_{r}\right)\mid_{r_{0}} & =\frac{b'\left(r_{0}\right)-1}{\kappa r_{0}^{2}}.\label{eq:17}
\end{align}
Since $b'\left(r_{0}\right)<1$, by the flare-out condition, $\rho+p_{r}<0$
in the throat by the Eq.~(\ref{eq:17}). In the non-conservative
approach to unimodular gravity, this property still holds true. Therefore, independently of the non-conservative sector, traversable wormholes in $\mathrm{NUG}$ necessarily violate the radial NEC at the throat.

Although the non-conservation equation (\ref{eq:8}) does not generate new dynamic information about the wormhole geometry, it preserves a fundamental characteristic of the theory: the absence of an algebraic relationship between the Ricci scalar $R$ and the trace of the energy-momentum tensor $T$. This behavior contrasts with General Relativity ($\mathrm{GR}$), in which $R=-\kappa T$, and also with conservative $\mathrm{UG}$, where the usual conservation of the energy-momentum tensor leads to the integrated relation 
\begin{equation}
R+\kappa T=4\Lambda_{U},
\end{equation}
with $\Lambda_{U}$ a constant of integration. In $\mathrm{NUG}$, none of these restrictions are present, so the algebraic relation between $R$ and $T$ remains undetermined by the field equations, which we will discuss next.

\subsection{Curvature Reconstruction Method}

The underdetermined nature of the field equations suggests an alternative strategy for constructing wormhole solutions in $\mathrm{NUG}$. Since no algebraic relation between $R$ and $T$ is provided by the theory, we treat the Ricci scalar as an independent geometric quantity and prescribe a phenomenologically motivated curvature profile $R\left(r \right)$. 

For a given redshift function $\Phi\left(r\right)$, the geometric definition of the Ricci scalar, Eq.~(\ref{eq:16-1}), becomes a differential equation for the shape function $b\left(r\right)$,
\begin{align}
b'\left(r\right)&=\frac{r^{2}}{2+\Phi'r}\left\{ R\left(r\right)+2\left(1-\frac{b}{r}\right)\Phi'^{2}+\frac{4}{r}\left(1-\frac{3}{4}\frac{b}{r}\right)\Phi'+2\left(1-\frac{b}{r}\right)\Phi''\right\}.\label{eq:Rr} 
\end{align}
Therefore, specifying the pair $\{R\left(r\right),\Phi\left(r\right)\}$ completely determines the wormhole geometry through the reconstruction of the shape function $b\left(r\right)$. Once $b\left(r\right)$ is obtained, the combinations $\left(\rho+p_r\right)$ and $\left(\rho+p_t\right)$ follow directly from Eqs.~(\ref{eq:15})-(\ref{eq:16}).

It should be emphasized that the reconstruction equations are identical to those obtained from the trace-free field equations. Their interpretation as a curvature-reconstruction scheme, however, is only consistent within the non-conservative formulation, where no algebraic relation constrains the Ricci scalar through the matter trace. Consequently, this reconstruction strategey constitutes a distinctive feature of $\mathrm{NUG}$, since neither $\mathrm{GR}$ nor conservative $\mathrm{UG}$ allow the Ricci scalar to be prescribed independently of the trace of the energy-momentum tensor. In this framework, the Ricci scalar acts as the primary geometrical input for the spacetime reconstruction, while the pair $\{R(r),\Phi(r)\}$ completely specifies the reconstructed wormhole solution. Having established the general reconstruction procedure, we now apply it to a phenomenologically motivated curvature profile.

\section{Wormhole Geometries from a Power-Law Curvature Profile}\label{sec:4}
To illustrate the reconstruction procedure, we adopt a power-law profile for the Ricci scalar,
\begin{align}
R(r)=R_0\left(\frac{r_0}{r}\right)^n, \label{eq:R}
\end{align}
where $R_0$ sets the curvature scale at the throat, while the exponent $n>0$ controls the spatial localization of the curvature. Since $r\geq r_0>0$, the prescribed profile remains finite throughout the wormhole domain and decays as $R(r)\to0$ for $r\to\infty$. This family of profiles provides a simple yet sufficiently general parametrization, allowing us to investigate how different curvature distributions affect the reconstructed wormhole geometry while preserving full analytical control over the reconstruction procedure. Power-law profiles are widely employed in gravitational physics because they provide simple localized functions with a controlled asymptotic behavior. They also appear naturally in several astrophysical contexts, such as galactic mass distributions and dark matter halo models \cite{bookBinney, Navarro_1997}, providing an additional phenomenological motivation for considering this class of curvature profiles.

For each choice of the redshift function, we reconstruct the shape
function $b\left(r\right)$, analyze the flare-out conditions, and investigate
the behavior of the null-energy combinations $\left(\rho+p_r\right)$ and
$\left(\rho+p_t\right)$.

\subsection{Constant Redshift Function.}
We begin our analysis by considering the simplest and most physically relevant configuration, namely the zero-tidal-force wormhole. In the context of traversable wormholes, the absence of event horizons requires the redshift function $\Phi\left(r\right)$ to remain finite everywhere. A particularly simple realization of this condition is obtained by choosing a constant redshift function, $\Phi'\left(r\right)=0$, which also eliminates the tidal forces experienced by travelers crossing the wormhole \cite{Morris:1988cz}.

In this regime of zero tidal forces, the geometric differential equation for the scalar curvature, Eq.~(\ref{eq:Rr}), simplifies significantly to:
\begin{align}
R\left(r\right) = \frac{2b'\left(r\right)}{r^2}.\label{eq:difeq}
\end{align}
Substituting the prescribed power-law profile, Eq.~(\ref{eq:R}) into Eq.~(\ref{eq:difeq}) yields a direct first-order differential equation for the shape function
\begin{align}
b'\left(r\right) = \frac{R_0 r_0^n}{2} r^{2-n}.\label{eq:bprimePhi0}
\end{align}
By integrating Eq.~(\ref{eq:bprimePhi0}) and imposing the fundamental throat boundary condition $b\left(r_0\right) = r_0$, we obtain the exact analytical solution for the shape function. For any decay index $n \neq 3$, the geometry is given by:
\begin{align}
b\left(r\right) = r_0 - \frac{R_0 r_0^3}{2(3-n)} + \frac{R_0 r_0^n}{2(3-n)} r^{3-n}.\label{eq:bPhi0}
\end{align}
It is mathematically interesting to note that for the critical case $n=3$, the integration changes its branch, and the solution assumes a logarithmic profile, $b(r) = r_0 + \frac{R_0 r_0^3}{2} \ln(r/r_0)$. Although this case is mathematically admissible, we restrict the present analysis to $n\neq3$. Remarkably, for the zero-tidal-force configuration the reconstruction equation becomes completely independent of the redshift sector, and the wormhole geometry is determined solely by the prescribed curvature profile.

The flare-out condition ($b'\left(r_0\right)<1$) is given by
\begin{align}
R_{0}r_{0}^{2} & <2. \label{eq:flare}
\end{align} 
The flare-out condition establishes a geometric link
between the scalar curvature in the throat $R_{0}$ and the throat
radius $r_{0}$ of the wormhole; for instance, it implies that the
fundamental length scale is governed by $r_{0}$. This condition constrains
the dimensionless combination $R_{0}r_{0}^{2}$, implying that the
characteristic curvature at the throat cannot exceed the scale set
by the inverse squared throat radius.

Now, we can use the expression for $b\left(r\right)$, $b'\left(r\right)$,
$\Phi\left(r\right)$ and $R$ in the field equations (\ref{eq:15})
and (\ref{eq:16}), obtaining
\begin{align}
\kappa\left(\rho+p_{r}\right)&=\left(\frac{2-n}{2\left(3-n\right)}\right)\frac{R_{0}r_{0}^{n}}{r^{n}}-\left(1-\frac{R_{0}r_{0}^{2}}{2\left(3-n\right)}\right)\frac{r_{0}}{r^{3}},\label{eq:rhopr}\\ \kappa\left(\rho+p_{t}\right)&=\left(\frac{4-n}{4\left(3-n\right)}\right)\frac{R_{0}r_{0}^{n}}{r^{n}}+\left(\frac{1}{2}-\frac{R_{0}r_{0}^{2}}{4\left(3-n\right)}\right)\frac{r_{0}}{r^{3}}.
\label{eq:rhopt}
\end{align}
It is important to emphasize that the trace-free field equations determine
only the combinations $\left(\rho+p_{r}\right)$ and $\left(\rho+p_{t}\right)$.
The individual quantities $\rho$, $p_{r}$ and $p_{t}$ remain undetermined
unless an additional physical condition, such as, for example, an
equation of state, is imposed. This degeneracy is a direct consequence
of the absence of a fundamental relation between $R$ and $T$ in
the non-conservative unimodular framework. On the other hand, we can evaluate the conditions $\kappa\left(\rho+p_{r}\right)$ and $\kappa\left(\rho+p_{t}\right)$ in the throat by obtaining
\begin{align}
\kappa\left(\rho+p_{r}\right)|_{r_{0}}&=-\frac{1}{r_{0}^{2}}\left(1-\frac{R_{0}r_{0}^{2}}{2}\right) \label{eq:rhpprr0}\\\kappa\left(\rho+p_{t}\right)|_{r_{0}}&=\frac{1}{2r_{0}^{2}}\left(1+\frac{R_{0}r_{0}^{2}}{2}\right),
\label{eq:rhoptr0}
\end{align}
which is independent of the parameter $n$. 

While the flare-out condition constrains the geometry in the vicinity of the throat, the asymptotic behavior of the solution determines whether the reconstructed spacetime approaches a physically acceptable external geometry. Therefore, to determine the asymptotic structure of the reconstructed spacetime, we examine the large-distance limit of the shape function (\ref{eq:bPhi0}) and the corresponding behavior of the metric coefficients. This analysis allows us to identify the conditions under which the geometry becomes asymptotically flat or retains a residual curvature-induced deformation at large distances. The expression for $b(r)/r$ is given by
\begin{align}
\frac{b\left(r\right)}{r}&=\left(1-\frac{R_{0}r_{0}^{2}}{2\left(3-n\right)}\right)\frac{r_{0}}{r}+\frac{R_{0}r_{0}^{n}}{2\left(3-n\right)}r^{2-n}. \label{eq:b/rPhi0}
\end{align}
Investigating the asymptotic limit of the Eq.~(\ref{eq:b/rPhi0}), for $n>2$ and $n\neq 3$ we obtain
\begin{align}
\lim_{r\rightarrow\infty}&\frac{b}{r}=0,
\end{align}
i.e., asymptotically flat behavior. Now, for $n=2$ the asymtotic behavior becomes
\begin{align}
\lim_{r\rightarrow\infty}&\frac{b}{r}=\frac{R_{0}r_{0}^{2}}{2},
\end{align}
a constant less than one because of the flare-out condition $R_0r_0^2<2$. For $n<2$, the second term in Eq.~(\ref{eq:b/rPhi0}) grows with $r^{2-n}$ and therefore dominates the asymtotic behavior. This case,
\begin{align}
\lim_{r\rightarrow\infty}&\frac{b}{r}=+\infty,
\end{align}
this diverges leading us to disregard this case in our subsequent analysis.

Having established the asymptotic behavior of the shape function, we will now examine the asymptotic behavior of the effective fluid combinations $\kappa\left(\rho+p_{r}\right)$ and $\kappa\left(\rho+p_{t}\right)$, Eqs.~(\ref{eq:rhopr}) and (\ref{eq:rhopt}). The asymptotic behavior of these combinations is organized into two distinct classes: 
\begin{itemize}
\item The term $r^{-n}$ representing the contribution of the curvature scalar;
\item The $r^{-3}$ term is a mixed contribution of the curvature scalar and the asymptotic behavior induced by the
constant redshift function. 
\end{itemize}
For $n>3$, the curvature term $r^{-n}$ decays faster than the term $r^{-3}$. Thus, the asymptotic state of the effective fluid is governed entirely by the dominant term
\begin{align}
\lim_{r\rightarrow\infty}\kappa\left(\rho+p_{r}\right)\mid_{n>3}&\approx-\left(1-\frac{R_{0}r_{0}^{2}}{2\left(3-n\right)}\right)\frac{r_{0}}{r^{3}}.\label{eq:ass_necr}\\
\lim_{r\rightarrow\infty}\kappa\left(\rho+p_{t}\right)\mid_{n>3}&\approx\left(\frac{1}{2}-\frac{R_{0}r_{0}^{2}}{4\left(3-n\right)}\right)\frac{r_{0}}{r^{3}}.\label{eq:ass_nect}
\end{align}
For $n=2$, the proportional term $r^{-n}$ goes to zero due to the numerator $(2-n)$ in Eq.~(\ref{eq:rhopr}), and therefore the asymptotic state of the effective fluid is also governed by the term $r^{-3}$. On the other hand, for the combination $\kappa\left(\rho+p_{t}\right)$, the asymptotic state is governed by the term $r^{-n}$. Consequently, the asymptotic behavior of both cases in $n=2$ becomes
\begin{align}
\lim_{r\rightarrow\infty}\kappa\left(\rho+p_{r}\right)\mid_{n=2}&\approx-\left(1-\frac{R_{0}r_{0}^{2}}{2}\right)\frac{r_{0}}{r^{3}} \label{eq:ass_necrn2}\\
\lim_{r\rightarrow\infty}\kappa\left(\rho+p_{t}\right)\mid_{n=2}&\approx\left(\frac{R_{0}r_{0}^{2}}{2}\right)\frac{1}{r^{2}}. \label{eq:ass_nectn2}
\end{align}
 
\subsection{Logarithmic Redshift Function.}\label{subsec:2}
The logarithmic redshift function is motivated by the weak-field description of astrophysical systems exhibiting approximately flat rotation curves. For static spherically symmetric spacetimes, the temporal component of the metric is given by $g_{tt}=-e^{2\Phi(r)}$. In the weak-field regime, where $|\Phi|\ll1$, the metric can be expanded as $g_{tt}\simeq-(1+2\Phi)$, while the Newtonian limit is described by $g_{tt}\simeq-(1+2\Psi)$, with $\Psi(r)$ denoting the Newtonian gravitational potential \cite{Misner:1973prb, Carroll:2004st, Weinberg}. Consequently, the redshift function becomes directly identified with the Newtonian potential, $\Phi(r)\simeq\Psi(r)$.

For gravitational systems displaying nearly constant circular velocities, such as the singular isothermal sphere, the Newtonian potential assumes the logarithmic form \cite{Sparke_Gallagher_2007, bookBinney, Begeman:1991iy, 91143000}
\begin{align}
\Psi(r)=v_c^2\ln\left(\frac{r}{r_0}\right),
\end{align}
where $v_c$ denotes the asymptotic circular velocity. This correspondence naturally motivates the logarithmic redshift function adopted in the present reconstruction,
\begin{align*}
\Phi\left(r\right) & =v_{0}^{2}\ln\left(\frac{r}{r_{0}}\right),
\end{align*}
where $v_0=v_c/c$ is the dimensionless asymptotic circular velocity. Since observed galactic rotation velocities satisfy $v_c\ll c$, one has $v_0^2\ll1$ \cite{bookBinney, Sofue_2001}. Substituting these expression into the geometric
definition of the Ricci scalar, Eq.~(\ref{eq:Rr}) using the power-law of $R(r)$, Eq.~(\ref{eq:R}), yields a first-order
differential equation for the shape function $b\left(r\right)$, 
\begin{align}
b'\left(r\right)+\frac{\gamma}{r}b\left(r\right)&=\eta r^{2-n}+\xi, \label{eq:difb}
\end{align}
from which the wormhole geometry can be obtained analytically as
\begin{align}
b\left(r\right) & =\frac{\xi}{\gamma+1}r+\frac{\eta}{3+\gamma-n}r^{3-n}+\left[1-\frac{\eta r_{0}^{2-n}}{3+\gamma-n}-\frac{\xi}{\gamma+1}\right]\frac{r_{0}^{\gamma+1}}{r^{\gamma}},\label{eq:43.7}
\end{align}
where we have defined $\gamma\equiv\frac{v_{0}^{2}\left(2v_{0}^{2}+1\right)}{2+v_{0}^{2}}$,
$\xi\equiv\frac{2v_{0}^{2}\left(1+v_{0}^{2}\right)}{2+v_{0}^{2}}$
and $\eta\equiv\frac{R_{0}r_{0}^{n}}{2+v_{0}^{2}}$. The Eq.~(\ref{eq:43.7}) satisfies the throat condition $b\left(r_{0}\right)=r_{0}$.
The derivative of Eq.~(\ref{eq:43.7}) becomes
\begin{align}
b'\left(r\right) & =\frac{\xi}{\gamma+1}+\frac{\left(3-n\right)\eta}{3+\gamma-n}r^{2-n}-\gamma\left[1-\frac{\eta r_{0}^{2-n}}{3+\gamma-n}-\frac{\xi}{\gamma+1}\right]\frac{r_{0}^{\gamma+1}}{r^{\gamma+1}}.\label{eq:44.7}
\end{align}
At the throat, we obtain
\begin{align}
b'\left(r_{0}\right) & =\xi+\eta r_{0}^{2-n}-\gamma.\label{eq:45}
\end{align}
The flare-out condition $b'\left(r_{0}\right)<1$ leads to
\begin{align*}
\eta r_{0}^{2-n} & <1+\gamma-\xi
\end{align*}
Using the definitions of the $\eta$, $\xi$ and $\gamma$, one finds
\begin{align*}
\frac{R_{0}r_{0}^{2}}{2+v_{0}^{2}} & <1-\frac{v_{0}^{2}}{2+v_{0}^{2}}.
\end{align*}
and therefore
\begin{align*}
\left(2+v_{0}^{2}\right)\left(1+\gamma-\xi\right)=\left(2+v_{0}^{2}\right)\left(1-\frac{v_{0}^{2}}{2+v_{0}^{2}}\right) & =2.
\end{align*}
Consequently,
\begin{align}
R_{0}r_{0}^{2} & <2\label{eq:45.7}
\end{align}
is the only restriction imposed by the flare-out condition on the prescribed Ricci scalar. Remarkably, the final constraint is independent of
the velocity parameter $v_{0}$, depending only on the dimensionless
combination $R_{0}r_{0}^{2}$. In this case, the Eq.~(\ref{eq:45.7}) presents a universality characteristic of the flare-out condition in a geometric reconstruction system via scalar curvature $R$ prescribed as a power law in $\mathrm{NUG}$.

Now, we can use the expression for $b\left(r\right)$, $b'\left(r\right)$,
$\Phi\left(r\right)$ and $R$ in the field equations (\ref{eq:15})
and (\ref{eq:16}), obtaining
\begin{align}
\kappa\left(\rho+p_{r}\right) & =\frac{4v_{0}^{2}}{\left(2+v_{0}^{2}\right)\left(\gamma+1\right)}r^{-2}+\frac{\eta}{3+\gamma-n}\left(2-n-2v_{0}^{2}\right)r^{-n}-\nonumber \\
 & -\left(2v_{0}^{2}+1+\gamma\right)\left[1-\frac{\eta r_{0}^{2-n}}{3+\gamma-n}-\frac{\xi}{\gamma+1}\right]\frac{r_{0}^{\gamma+1}}{r^{\gamma+3}},\label{eq:46.7}\\
\kappa\left(\rho+p_{t}\right) & =\left(v_{0}^{2}-\frac{3}{2}\gamma+\frac{1}{2}\right)\left[1-\frac{\eta r_{0}^{2-n}}{3+\gamma-n}-\frac{\xi}{\gamma+1}\right]\frac{r_{0}^{\gamma+1}}{r^{\gamma+3}}\nonumber \\
 & +\frac{\eta r^{-n}}{2\left(3+\gamma-n\right)}\left[n\left(v_{0}^{2}-1\right)-v_{0}^{2}\left(1+\gamma\right)+2\left(2-\gamma\right)\right]+\nonumber \\
 & +\frac{2v_{0}^{2}\left(2v_{0}^{2}+1\right)}{\left(2+v_{0}^{2}\right)\left(\gamma+1\right)}r^{-2}.\label{eq:47.7}
\end{align}
We can evaluate the conditions $\kappa\left(\rho+p_{r}\right)$ and $\kappa\left(\rho+p_{t}\right)$ in the throat by obtaining
\begin{align}
\kappa\left(\rho+p_{r}\right)\mid_{r_{0}} & =-\frac{2}{\left(2+v_{0}^{2}\right)r_{0}^{2}}\left[1-\frac{R_{0}r_{0}^{2}}{2}\right]\label{eq:49.7}\\
\kappa\left(\rho+p_{t}\right)\mid_{r_{0}} & =\frac{1}{\left(2+v_{0}^{2}\right)r_{0}^{2}}\left[2v_{0}^{2}+1+\frac{R_{0}r_{0}^{2}\left(1-v_{0}^{2}\right)}{2}\right],\label{eq:50.7}
\end{align}
where the expressions are strictly independent of the parameter $n$.
Furthermore, the radial energy condition (\ref{eq:49.7}) is violated
at the throat (as we would expect from the structure of the field
equations in $\mathrm{NUG}$), while the tangential energy condition
assumes positive values, since the sign in brackets, in the Eq.~(\ref{eq:50.7}), is always positive since $v_{0}^{2}\ll 1$.

For $v_{0}=0$ $\left(\Phi'=0\right)$
and $n=4$; the reconstruted family reduces to
\begin{align}
b\left(r\right) & =-\frac{R_{0}r_{0}^{4}}{2}r^{-1}+\frac{R_{0}r_{0}^{3}}{2}+r_{0},\\
b'\left(r\right) & =\frac{R_{0}r_{0}^{4}}{2}r^{-2}.
\end{align}
The corresponding flare-out condition remains identical to Eq.~(\ref{eq:45.7}). Interestingly, the Ellis-Bronnikov geometry is recovered as a particular member of this reconstructed family when the curvature amplitude satisfies
\begin{align}
R_{0}r_{0}^{2}=-2,
\end{align}
for which the shape function becomes
\begin{align}
b(r)=\frac{r_{0}^{2}}{r}.
\end{align}
Therefore, the Ellis-Bronnikov geometry is not obtained by prescribing the shape function itself. Instead, it emerges naturally as a particular member of the family reconstructed from the Ricci scalar profile, illustrating the geometric reconstruction philosophy adopted throughout this work. 

It is instructive to investigate the large-distance limit of the shape function (\ref{eq:43.7}) and the corresponding behavior of the metric coefficients. The expression of $b\left(r\right)/r$ is given by
\begin{align}
\frac{b\left(r\right)}{r}&=\frac{\xi}{\gamma+1}+\frac{\eta}{3+\gamma-n}r^{2-n}+\left[1-\frac{\eta r_{0}^{2-n}}{3+\gamma-n}-\frac{\xi}{\gamma+1}\right]\frac{r_{0}^{\gamma+1}}{r^{\gamma+1}}. \label{eq:b/r}
\end{align}
The asymptotic limit of the Eq.~(\ref{eq:b/r}) for $n>2$ is
\begin{align}
\lim_{r\rightarrow\infty}\frac{b\left(r\right)}{r}&=\frac{\xi}{\gamma+1}\approx v_{0}^{2}, \label{eq:b/rm2}
\end{align}
where we have used the expressions for $\xi$ and $\gamma$, together with the observationally motivated condition $v^{2}_{0}\ll 1$. Now, for $n=2$ the asymptotic behavior becomes
\begin{align}
\lim_{r\rightarrow\infty}\frac{b\left(r\right)}{r}&=\frac{\xi}{\gamma+1}+\frac{\eta}{\gamma+1}\approx v_{0}^{2}+\frac{R_{0}r_{0}^{2}}{2}, \label{eq:b/rn2}
\end{align}
where we have used the expressions for $\xi$, $\gamma$ and $\eta$, together with the observationally motivated condition $v_0\ll 1$. It is worth noting that in both cases, we have that $\lim_{r\rightarrow \infty}r^{-(\gamma+1)}=0$, since $\gamma>0$. For $n<2$, the second term in Eq.~(\ref{eq:b/r}) grows as $r^{2-n}$ and therefore dominates the asymptotic behavior. In this case,
\begin{align}
\lim_{r\rightarrow\infty}\frac{b(r)}{r}
=
\pm \infty ,
\end{align}
depending on the sign of the coefficient $\eta/(3+\gamma-n)$. Consequently, the condition $b(r)/r\rightarrow 0$ is not satisfied and the corresponding geometries do not possess a well-defined asymptotic region. For this reason, curvature profiles with $n<2$ will not be considered in the subsequent analysis. Physically relevant solutions are restricted to $n \geq 2$. While the cases with $n > 2$ exhibit a faster decay of the curvature sector, the $n=2$ scenario represents a critical configuration where the curvature contribution to the asymptotic geometry remains finite. It should be emphasized that, for $v_0\neq0$, the reconstructed geometries are not asymptotically flat in the strict Morris-Thorne sense, since $b(r)/r$ approaches a non-vanishing constant. This behavior is a direct consequence of the logarithmic redshift function and reflects the presence of an effective matter distribution extending to large distances.

Having established the asymptotic behavior of the shape function, we now examine the effective fluid combination. Using Eqs.~(\ref{eq:46.7})–(\ref{eq:47.7}), the asymptotic expansion shows that all contributions to the effective fluid $\left(\rho+p_r\right)$ and $\left(\rho+p_t\right)$ can be organized into three distinct decay classes: terms proportional to $r^{-2}$, terms proportional to $r^{-n}$, and a rapidly decaying term proportional to $r^{-(\gamma+3)}$. Physically, each class corresponds to a distinct geometric and material source:
\begin{itemize}
\item The $r^{-2}$ contributions are associated with the asymptotic behavior induced by the logarithmic redshift function.
\item The $r^{-n}$ terms encode the direct dynamical contribution of the prescribed scalar curvature;
\item And finally, the $r^{-(\gamma+3)}$ terms represent the pure wormhole geometric footprint. Since $\gamma > 0$, this specific contribution decays faster than the background mass distribution, ensuring that the exotic matter properties required to keep the throat open are strongly localized near $r_0$ and become completely negligible at large distances.
\end{itemize}

To make this asymptotic behavior explicit, let us evaluate the large-distance limit of the effective fluid sector. For $n>2$, the curvature terms $r^{-n}$ decay faster than the term induced by logarithmic redshift function. Thus, the asymptotic state of the fluid is governed entirely by the $r^{-2}$ contributions:
\begin{align}
\lim_{r\rightarrow\infty}\kappa\left(\rho+p_{r}\right) & \approx \frac{4v_{0}^{2}}{\left(2+v_{0}^{2}\right)\left(\gamma+1\right)}r^{-2},\\
\lim_{r\rightarrow\infty}\kappa\left(\rho+p_{t}\right) & \approx \frac{2v_{0}^{2}\left(2v_{0}^{2}+1\right)}{\left(2+v_{0}^{2}\right)\left(\gamma+1\right)}r^{-2}.
\end{align}
In this regime, the wormhole geometry becomes effectively localized around the throat, while the asymptotic spacetime is entirely governed by the logarithmic redshift function. Conversely, for the critical case $n=2$, the scalar-curvature contribution exhibits the same asymptotic radial scaling as the logarithmic redshift function. Consequently, the curvature continues to influence the spacetime at arbitrarily large distances, and the asymptotic expressions become
\begin{align}
\lim_{r\rightarrow\infty}\kappa\left(\rho+p_{r}\right) & \approx 2v_0^{2}\left[1- \frac{R_0 r_0^{2}}{2}\right] r^{-2},\\
\lim_{r\rightarrow\infty}\kappa\left(\rho+p_{t}\right) & \approx \left[v_0^{2}+ \frac{R_0 r_0^{2}}{2}\right] r^{-2}.
\end{align}
In this scenario, the prescribed curvature profile leaves a persistent imprint on the asymptotic behavior of the effective quantities $\kappa\left(\rho+p_r\right)$ and $\kappa\left(\rho+p_t\right)$, so that the logarithmic redshift function continues to influence the reconstructed solution even at arbitrarily large distances.

\subsection{Inverse-Radial Redshift Function.}
Unlike the previous choices, the inverse-radial redshift function introduces a non-trivial gravitational potential while remaining everywhere finite and asymptotically vanishing. Consequently, the spacetime is free of horizons and smoothly approaches the weak-field regime at large distances. Moreover, this profile provides a simple analytical framework to investigate how a localized gravitational potential modifies the reconstructed wormhole geometry in $\mathrm{NUG}$. Therefore, we adopt the redshift function
\begin{align*}
\Phi\left(r\right) & =\frac{r_0}{r}.
\end{align*}
Substituting these expression into the geometric
definition of the Ricci scalar, Eq.~(\ref{eq:Rr}) using the power-law of $R(r)$, Eq.~(\ref{eq:R}), yields a first-order
differential equation for the shape function $b\left(r\right)$, 
\begin{align}
b'\left(r\right)+\left(\frac{2r_{0}^{2}}{\left(2r-r_{0}\right)r^{2}}+\frac{r_{0}}{\left(2r-r_{0}\right)r}\right)b\left(r\right)&=\frac{R_{0}r_{0}^{n}}{\left(2r-r_{0}\right)}r^{3-n}+\frac{2r_{0}^{2}}{\left(2r-r_{0}\right)r}, \label{eq:difb2}
\end{align}
from which the wormhole geometry can be obtained as
\begin{align}
b\left(r\right)&=\frac{e^{-\frac{2r_{0}}{r}}r^{5}}{\left(2r-r_{0}\right)^{5}}\left[e^{2}b_{0}+R_{0}r_{0}^{n}\int_{r_{0}}^{r}\frac{\left(2r-r_{0}\right)^{4}e^{\frac{2r_{0}}{r}}}{r^{2+n}}dr+2r_{0}^{2}\int_{r_{0}}^{r}\frac{\left(2r-r_{0}\right)^{4}e^{\frac{2r_{0}}{r}}}{r^{6}}dr\right].\label{eq:brPhir0r}
\end{align}
The last integral on the right-hand side of Eq.~(\ref{eq:brPhir0r}) can be solved by applying successive integration by parts. Evaluating the integrals and reorganizing the terms, we arrive at the closed-form analytical solution:
\begin{align}
b\left(\epsilon\right) &= \frac{e^{-2\epsilon}}{\left(2-\epsilon\right)^{5}}\left(b_{0}+\frac{21}{2}r_{0}\right)e^{2} \nonumber \\
&\quad - \frac{r_{0}}{\left(2-\epsilon\right)^{5}}\left[\left(2-\epsilon\right)^{4}+2\left(2-\epsilon\right)^{3}+3\left(2-\epsilon\right)^{2}+3\left(2-\epsilon\right)+\frac{3}{2}\right] \nonumber \\
&\quad - \frac{e^{-2\epsilon}R_{0}r_{0}^{3}}{\left(2-\epsilon\right)^{5}}\left[J_{n}\left(\epsilon\right)-8J_{n-1}\left(\epsilon\right)+24J_{n-2}\left(\epsilon\right)-32J_{n-3}\left(\epsilon\right)+16J_{n-4}\left(\epsilon\right)\right] \label{eq:bxPhi}
\end{align}
where we have introduced the dimensionless inverse radial coordinate $\epsilon=\frac{r_0}{r}$, and defined the integral operator:
\begin{align}
J_{k}\left(\epsilon\right)&\equiv\int_{1}^{\epsilon}\bar{\epsilon}^{k}e^{2\bar{\epsilon}}d\bar{\epsilon}.
\end{align}
The analytical expression (\ref{eq:bxPhi}) elegantly encapsulates an entire family of wormhole geometries without the need to evaluate case-by-case integrations. The boundary condition at the throat, $b(r_0)=r_0$ (which corresponds to $\epsilon=1$), directly fixes the integration constant to $b_0=r_0$. 

The mathematical nature of the geometry bifurcates depending strictly on the parameter $n$. For rapidly decaying curvature profiles ($n \geq 4$), the index $k$ in the integral operator is always non-negative. Consequently, $J_k(\epsilon)$ admits an elementary closed form involving polynomials multiplied by $e^{2\epsilon}$, so that the wormhole shape function is entirely described by rational functions and standard exponentials. Conversely, for slower curvature decays, such as the cases $n=2$ or $n=3$, negative exponents arise in the integral operator. In these cases, $J_k(\epsilon)$ generally involves the exponential integral function, $\operatorname{Ei}(2\epsilon)$, in addition to elementary terms.

Remarkably, the local physics at the throat can be determined without resorting to the full integrated expression. By evaluating the original differential equation (\ref{eq:difb2}) precisely at the throat $r = r_0$, and using $b(r_0)=r_0$, the terms simplify drastically to:
\begin{align}
b'\left(r_{0}\right) &= R_{0}r_{0}^{2}-1. \label{eq:bprime_throat_exp}
\end{align}
Applying the flare-out condition, $b'(r_0) < 1$, we obtain the constraint:
\begin{align}
R_{0}r_{0}^{2} < 2, \label{eq:flare_exp}
\end{align}
which reveals that the curvature scale bound at the throat is universally identical to the previous models, completely independent of the asymptotic decay parameter $n$. Unlike the cases studied previously, the function $b'(r)$ evaluated at the throat can assume negative values, as Eq.~(\ref{eq:bprime_throat_exp}) for $R_0 r_0^2<1$. This characteristic is strictly related to the form of the redshift function chosen.

Finally, substituting Eq.~(\ref{eq:bprime_throat_exp}) into the general combination $\kappa\left(\rho+p_r\right)$ and $\kappa\left(\rho+p_t\right)$ evaluated at the throat, Eqs.~(\ref{eq:15})-(\ref{eq:16}), we find:
\begin{align}
\kappa\left(\rho+p_{r}\right)\mid_{r_{0}} & = \frac{R_{0}r_{0}^{2}-2}{r_{0}^{2}}\label{eq:nec_exp} \\
\kappa\left(\rho+p_{t}\right)\mid_{r_{0}} & = \frac{R_{0}r_{0}^{2}-1}{r_{0}^{2}}.\label{eq:nect_exp} 
\end{align}
Imposing the flare-out condition (\ref{eq:flare_exp}), the numerator is strictly negative in the Eq.~(\ref{eq:nec_exp}). Therefore, $\kappa(\rho+p_r)\mid_{r_0} < 0$ for all $n$, confirming that the generic requirement for effective exotic matter at the throat persists under an exponential redshift configuration in $\mathrm{NUG}$. On the other hand, unlike the previous cases, the condition $\kappa\left(\rho+p_t\right)$ can assume negative values in the throat as long as the condition $R_0r_0^{2}<1$, note that this condition does not violate the flare-out condition.

While the analytical determination of the local constraints at the throat is straightforward, substituting the general shape function (\ref{eq:bxPhi}) into the trace-free field equations to extract the full spatial profiles of the effective matter sector ($\rho+p_r$, and $\rho+p_t$) yields exceedingly lengthy and unilluminating algebraic structures. Furthermore, analyzing the asymptotic flatness via the limit $\lim_{r\to\infty} b(r)/r$ using the integral operators $J_k(\epsilon)$ requires careful numerical tracking of the exponential integrals. 

Therefore, rather than displaying these cumbersome expressions, the global physical properties of the effective fluid distribution and the asymptotic behavior of the spacetime geometry for this exponential class are more effectively investigated graphically. A detailed numerical evaluation of the NEC and the metric profiles is presented in the following section, providing a clear visualization of how the exotic matter is localized and how the spacetime evolves towards spatial infinity.

\section{Numerical Analysis.}\label{sec:5}

To facilitate the numerical analysis and to highlight the dimensionless structure of the reconstructed solutions, it is convenient to introduce the dimensionless radial coordinate
\begin{align}
x=\frac{r}{r_0},
\end{align}
together with the dimensionless curvature parameter
\begin{align}
\lambda \equiv R_0 r_0^{2}.\label{eq:dimlambda}
\end{align}
Consequently, the dimensionless representation reveals a geometric similarity among reconstructed solutions sharing the same value of $\lambda=R_0 r_0^{2}$. Wormholes with different throat radii but identical values of $\lambda$ therefore belong to the same dimensionless geometric family, differing only by an overall rescaling of the length scale. For this reason, all numerical results presented below are expressed in terms of the dimensionless coordinate $x=r/r_0$. 

\subsection{Case of Constant Redshift Function, $\Phi\left(r\right)=\mathrm{cte}$.}

In terms of dimensionless variables, the reconstructed wormhole geometry can be expressed entirely in terms of the parameter set ($x$, $\lambda$, $n$), while the explicit dependence on the throat radius disappears from the dimensionless representation. As an illustration, the dimensionless form of the shape function, Eq. (\ref{eq:bPhi0}), becomes
\begin{align}
\frac{b\left(x\right)}{r_{0}}&=1-\frac{\lambda}{2\left(3-n\right)}+\frac{\lambda}{2\left(3-n\right)}x^{3-n}.\label{eq:b/r0x}
\end{align}

\begin{figure}[h!]
    \centering
    \includegraphics[width=0.8\textwidth]{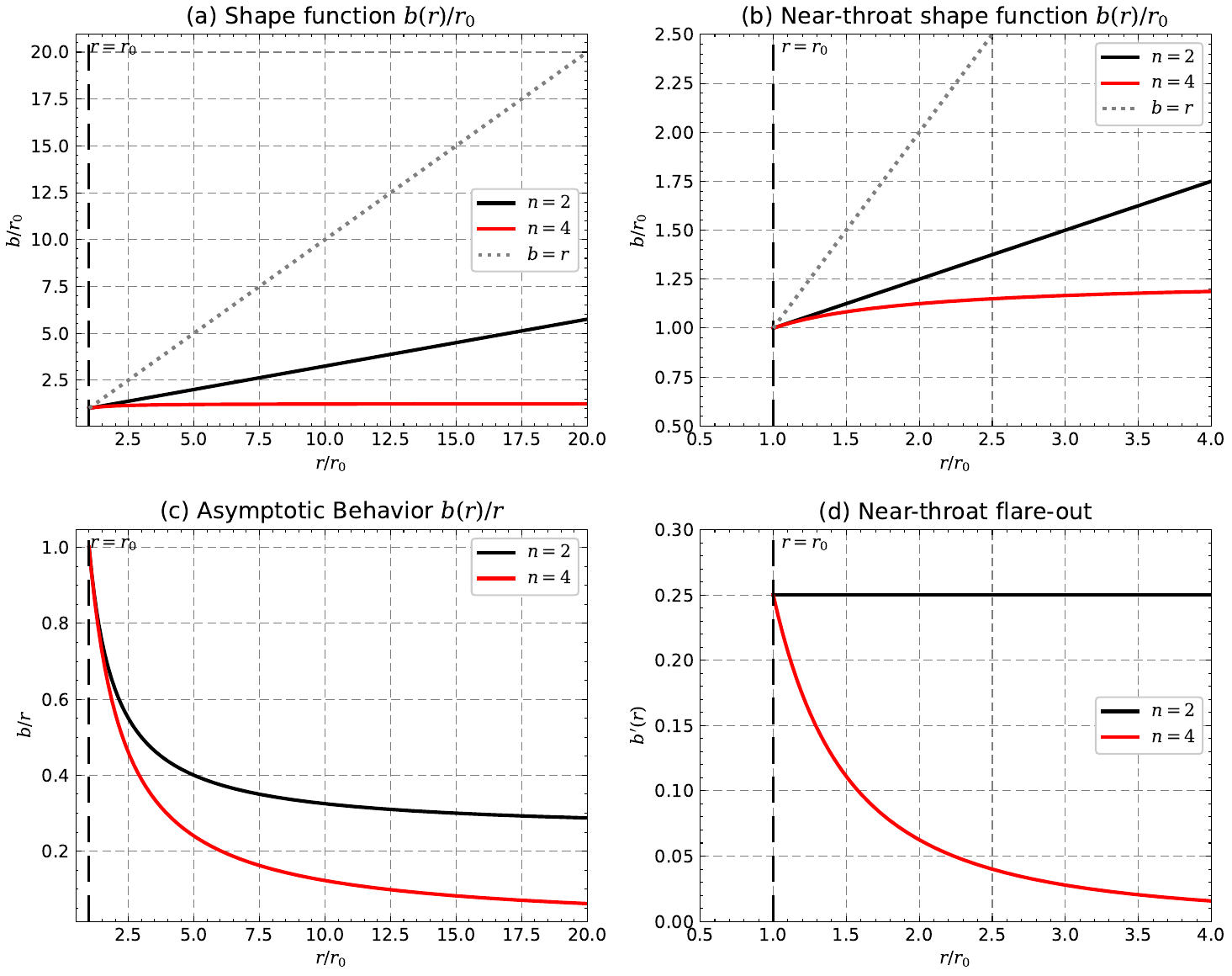}
    \caption{Geometric properties of the reconstructed wormhole supported by constant redshift function for $\lambda=0.5$ and different values of $n$. Panel (a) shows the shape function, panel (b) illustrates near-throat shape function, while panel (c) its asymptotic behavior through the ratio $b(r)/r$, (d) illustrates near-throat flare-out condition.}
    \label{fig:minha-imagem4}
\end{figure}

\begin{figure}[h!]
\centering
    \includegraphics[width=0.8\textwidth]{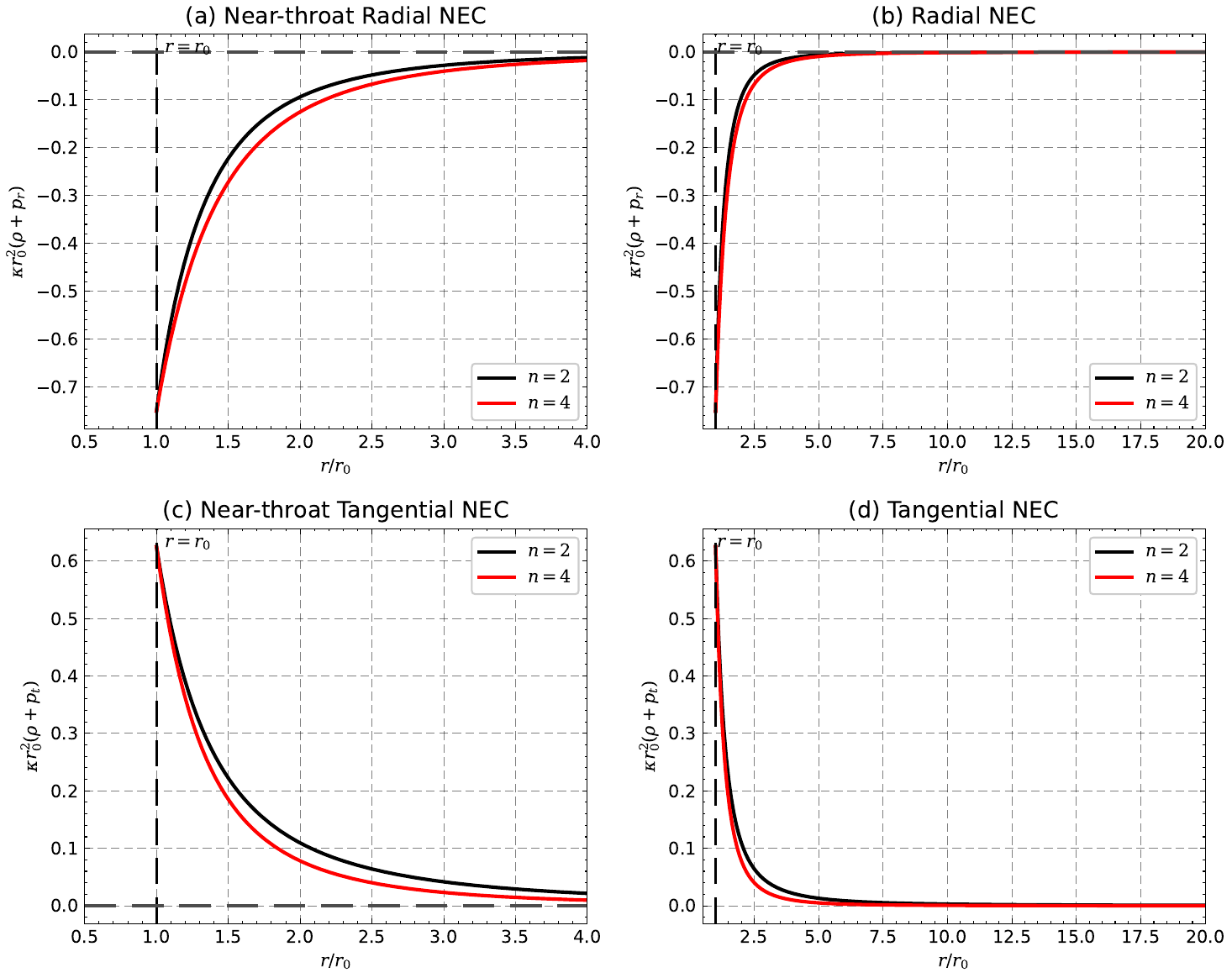}
    \caption{Radial and tangential null-energy conditions for the case of a constant redshift function, $\Phi' =0$. The upper panels show the dimensionless radial NEC combination, $\kappa r_0^{2}\left(\rho+p_r\right)$, with the left panel providing a magnified view near the throat and the right panel displaying the behavior up to $r=20r_0$. The lower panels present the corresponding dimensionless tangential NEC combination, $\kappa r_0^{2}\left(\rho+p_t\right)$, with the same radial ranges. The curvature parameter is fixed at $\lambda=0.5$, while two values of the prescribed curvature exponent are considered, namely $n=2$ and $n=4$, illustrating the influence of the curvature profile on the energy conditions.}
    \label{fig:minha-imagem5}
\end{figure}

\begin{figure}[htbp]
  \centering
  \begin{subfigure}[b]{0.49\textwidth}
    \includegraphics[width=\linewidth]{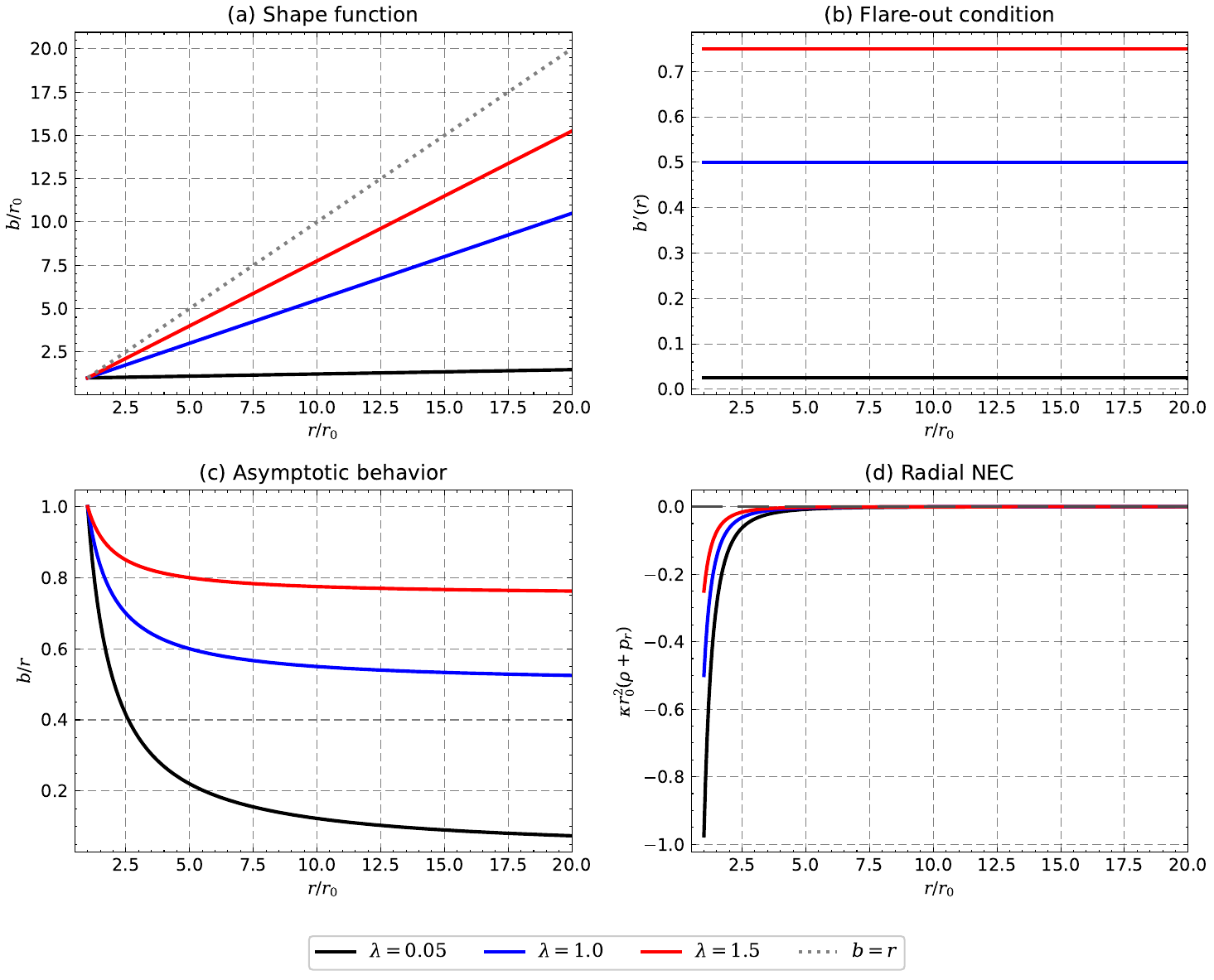}
    \label{fig:figura1}
  \end{subfigure}
   \hfill
  \begin{subfigure}[b]{0.49\textwidth}
    \includegraphics[width=\linewidth]{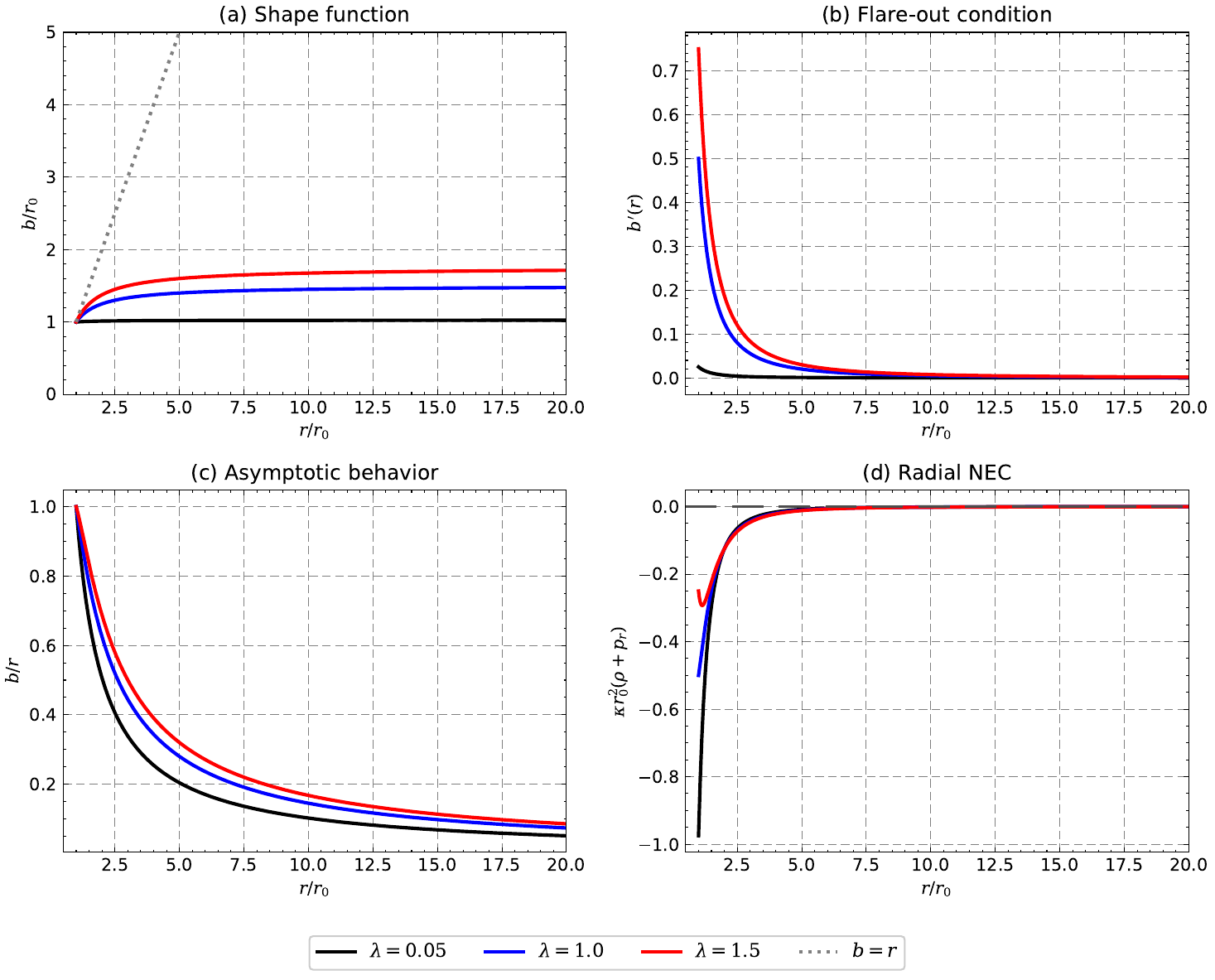}
    \label{fig:figura2}
  \end{subfigure}
  \caption{Geometrical and physical properties of the reconstructed wormhole supported by constant redshift function for fixed $n=2$ (Left Panel), $n=4$ (Right Panel) and different values of the dimensionless curvature parameter $\lambda$. Panels (a)–(c) illustrate the influence of the curvature amplitude on the wormhole geometry, while panel (d) shows the corresponding behavior of the radial null energy condition. Increasing $\lambda$ produces a wider geometry and progressively reduces the amount of exotic matter required near the throat.}
  \label{fig:duas_figuras}
\end{figure}

The numerical behavior of the reconstructed solutions is illustrated in Figs. (\ref{fig:minha-imagem4})–(\ref{fig:duas_figuras}). We first analyze the influence of the curvature exponent $n$, followed by the dependence on the dimensionless curvature parameter $\lambda$.

\textbf{Figure}~(\ref{fig:minha-imagem4}) presents the geometrical properties of the reconstructed wormhole for different values of the curvature exponent $n$. In all cases, the throat condition $b(r_0)=r_0$ is automatically satisfied by construction, while the flare-out condition remains fulfilled since $b'(r_0)<1$. 
\begin{itemize}
\item The shape function $b(r)/r_0$, displayed in panel (a), exhibits a monotonic increase from the throat while preserving the required wormhole behavior. As the curvature exponent $n$ increases, the reconstructed shape function approaches its asymptotic regime more rapidly. This behavior is expected since the curvature profile $R(r)$, Eq. (\ref{eq:R}), decays faster for larger values of $n$, reducing the contribution of the curvature source away from the throat. This feature becomes more evident in the near-throat region shown in panel (b), where the differences between the reconstructed solutions are amplified. Although all curves satisfy the same throat condition, increasing $n$ produces a steeper transition immediately outside the throat, indicating that the curvature distribution becomes progressively more localized around the wormhole throat.
\item The asymptotic behavior, illustrated in panel (c), reveals that the reconstructed solution with $n=2$ does not satisfy the asymptotic flatness condition, since the ratio $b(r)/r$ approaches a non-vanishing constant at large distances. This behavior reflects the slower decay of the curvature profile, allowing the curvature contribution to remain relevant even far from the throat. In contrast, for larger values of the curvature exponent, represented here by $n=4$, the ratio $b(r)/r$ rapidly approaches zero, indicating that the reconstructed geometry becomes asymptotically flat. 
\item Finally, panel (d) presents a magnified view of the derivative of the shape function in the vicinity of the throat. As expected, all reconstructed solutions satisfy the flare-out condition, with $b'(r_0)<1$. The distinct behavior exhibited by the different curves follows directly from the analytical expression obtained in the previous section. In particular, for $n=2$ the derivative of the shape function remains constant, whereas for larger values of the curvature exponent it decreases as a power law with the radial coordinate. Although only the near-throat region is displayed, this behavior persists throughout the entire spacetime. Therefore, increasing the curvature exponent progressively confines the flaring-out behavior to the vicinity of the throat, consistently with the faster decay of the curvature profile and the asymptotic behavior discussed above.
\end{itemize}

\textbf{Figure}~\ref{fig:minha-imagem5} illustrates the behavior of the null energy conditions for the reconstructed wormhole solutions. Since traversable wormholes necessarily require violations of the classical energy conditions within $\mathrm{GR}$, the radial and tangential null energy conditions provide a direct measure of the amount and distribution of exotic matter supporting the geometry. 
\begin{itemize}
\item Panels (a) and (b) display the radial NEC, both in the vicinity of the throat and over the entire integration domain. In all reconstructed solutions the radial NEC is violated, confirming the presence of the exotic matter required to sustain the wormhole geometry. The different behaviors observed for the curvature exponent $n$ follow directly from Eq.~(\ref{eq:rhopr}). In particular, for $n=2$ the curvature contribution proportional to $r^{-n}$ vanishes identically, leaving the radial NEC entirely governed by the geometric term proportional to $r^{-3}$. For larger values of $n$, the curvature sector contributes near the throat but decays progressively faster with increasing radial distance. Consequently, its compensating effect on the negative geometrical contribution becomes weaker, leading to a more pronounced violation of the radial NEC as the curvature exponent increases.
\item Panels (c) and (d) display the tangential NEC, which remains positive throughout the entire integration domain for all reconstructed solutions considered here. Consequently, no violation of the tangential NEC is observed. As in the radial sector, the behavior follows directly from the analytical solution Eq. (\ref{eq:rhopt}). For $n=2$, the tangential NEC receives contributions from both the curvature and geometric terms, whereas for $n=4$ the curvature contribution $r^{-n}$ vanishes identically, leaving the solution entirely governed by the geometric term proportional to $r^{-3}$. This explains the faster decay observed for larger values of the curvature exponent, while preserving the positivity of the tangential NEC.
\end{itemize}

\textbf{Figure}~\ref{fig:duas_figuras} illustrates the influence of the dimensionless curvature parameter $\lambda=R_0r_0^2$ on the reconstructed wormhole geometry for fixed values of the curvature exponent, namely $n=2$ (left panels) and $n=4$ (right panels). Since $\lambda$ measures the amplitude of the curvature profile, varying this parameter allows one to investigate how the geometry and the corresponding energy conditions respond to changes in the curvature scale while preserving the same functional form of the reconstruction.
\begin{itemize}
\item Panels (a) show that increasing the dimensionless curvature parameter produces a systematic enlargement of the reconstructed wormhole geometry. Although all solutions satisfy the throat condition $b(r_0)=r_0$, larger values of $\lambda$ lead to a steeper growth of the shape function immediately outside the throat. This behavior reflects the stronger contribution of the curvature sector to the reconstructed geometry as the amplitude of the curvature profile increases.
\item Panels (b) illustrate the behavior of the derivative of the shape function for different values of the dimensionless curvature parameter. Two qualitatively distinct regimes emerge depending on the curvature exponent. For $n=2$, shown in the left panel, the derivative of the shape function remains constant throughout the radial domain, so that increasing $\lambda$ simply shifts the constant value of $b'(r)$ while preserving the flare-out condition. In contrast, for $n=4$, displayed in the right panel, the derivative decreases rapidly away from the throat, reflecting the stronger radial decay of the curvature profile. Although increasing $\lambda$ raises the overall magnitude of $b'(r)$, all solutions rapidly approach zero outside the near-throat region. In both cases, the flare-out condition remains satisfied provided $\lambda<2$, showing that the admissible values of the dimensionless curvature parameter are naturally constrained by the wormhole geometry.
\item Panels (c) illustrate the asymptotic behavior of the reconstructed solutions through the ratio $b(r)/r$. For both values of the curvature exponent, increasing the dimensionless curvature parameter modifies only the amplitude of the reconstructed geometry, without altering its asymptotic character. In particular, the solution with $n=2$ remains non-asymptotically flat for values $\lambda=1.0$ and $\lambda=1.5$, since the ratio $b(r)/r$ approaches a non-vanishing constant at large distances. On the other hand, for $n=4$ all reconstructed solutions satisfy the asymptotic flatness condition, with $b(r)/r$ rapidly approaching zero independently of the particular value of $\lambda$.
\item Finally, panels (d) show the behavior of the radial null energy condition for different values of the dimensionless curvature parameter. For both values of the curvature exponent, increasing $\lambda$ systematically reduces the magnitude of the NEC violation, causing the curves to approach the null value while remaining negative throughout the radial domain. Consequently, larger values of the curvature parameter decrease the amount of exotic matter required to sustain the wormhole geometry. Nevertheless, this improvement is naturally limited by the flare-out condition, which requires $\lambda<2$. Therefore, the dimensionless curvature parameter plays a dual role: it controls the geometrical deformation of the reconstructed wormhole while simultaneously regulating the amount of exotic matter necessary to support the solution.
\end{itemize}

Overall, the numerical analysis demonstrates that the curvature exponent $n$ determines the qualitative geometric class of the reconstructed wormhole, whereas the dimensionless curvature parameter $\lambda$ regulates the intensity of both the geometrical deformation and the violation of the radial null energy condition.

\subsection{Case of Logarithmic Redshift Function, $\Phi\left(r\right)= v_{0}^{2}\ln\left(\frac{r}{r_{0}}\right)$.}

In terms of dimensionless variables, the reconstructed wormhole geometry can be expressed entirely in terms of the parameter set ($x$, $\lambda$, $n$, $v_0$), while the explicit dependence on the throat radius disappears from the dimensionless representation.
As an illustration, the dimensionless form of the shape function, Eq.~(\ref{eq:43.7}), becomes
\begin{align}
\frac{b\left(x\right)}{r_{0}}&=\frac{\xi}{\gamma+1}x+\frac{\lambda}{\left(2+v_{0}^{2}\right)\left(3+\gamma-n\right)}x^{3-n}+\left[1-\frac{\lambda}{\left(2+v_{0}^{2}\right)\left(3+\gamma-n\right)}-\frac{\xi}{\gamma+1}\right]x^{-\gamma}. \label{eq:bxadmi}
\end{align}

\begin{figure}[t!]
    \centering
    \includegraphics[width=0.8\textwidth]{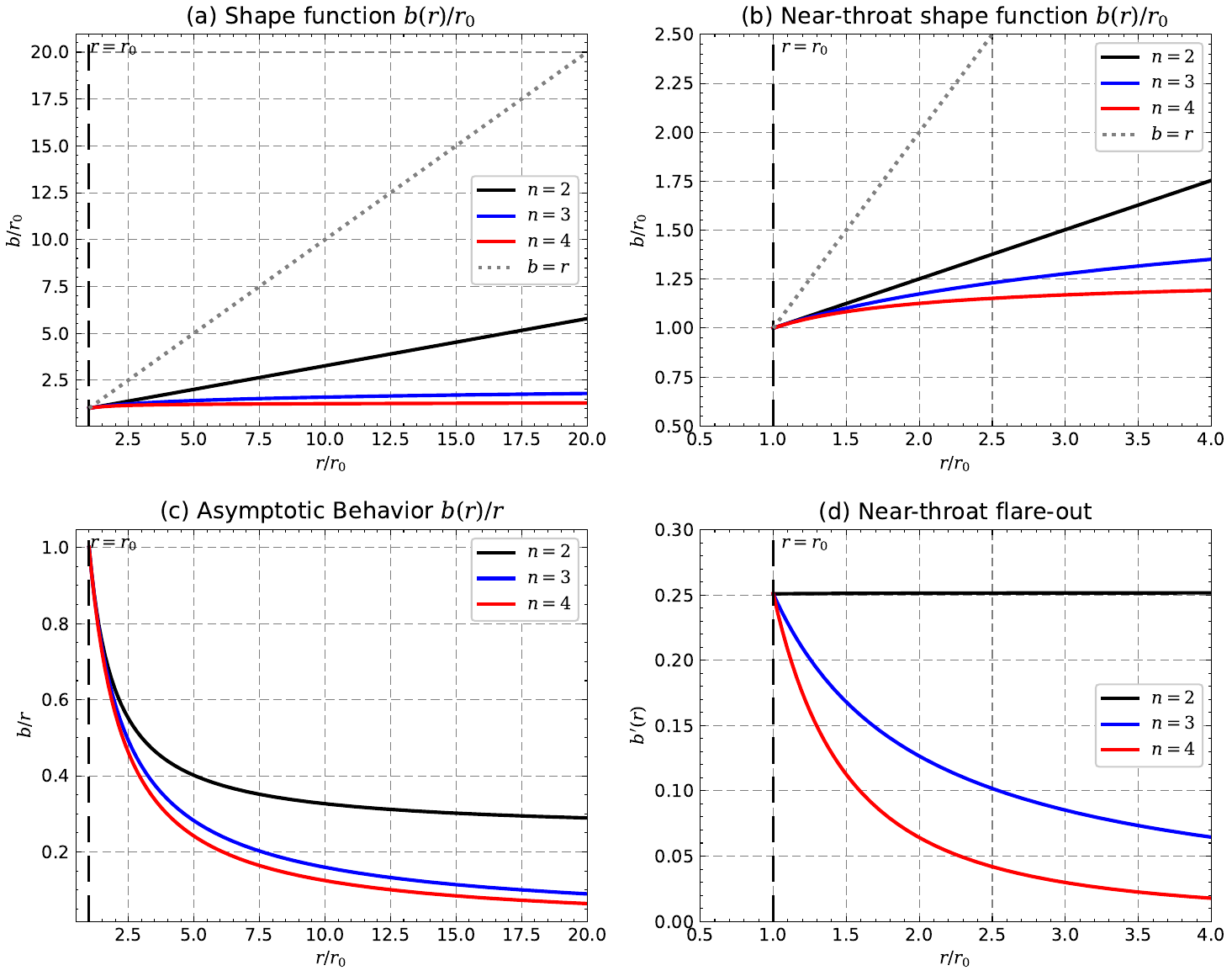}
    \caption{Geometric properties of the reconstructed wormhole supported by logarithmic redshift function for $\lambda=0.5$, $v_0=0.05$ and different values of $n$. Panel (a) shows the shape function, panel (b) illustrates near-throat shape function, while panel (c) its asymptotic behavior through the ratio $b(r)/r$, (d) illustrates near-throat flare-out condition.}
    \label{fig:minha-imagem}
\end{figure}

\begin{figure}[h!]
\centering
    \includegraphics[width=0.8\textwidth]{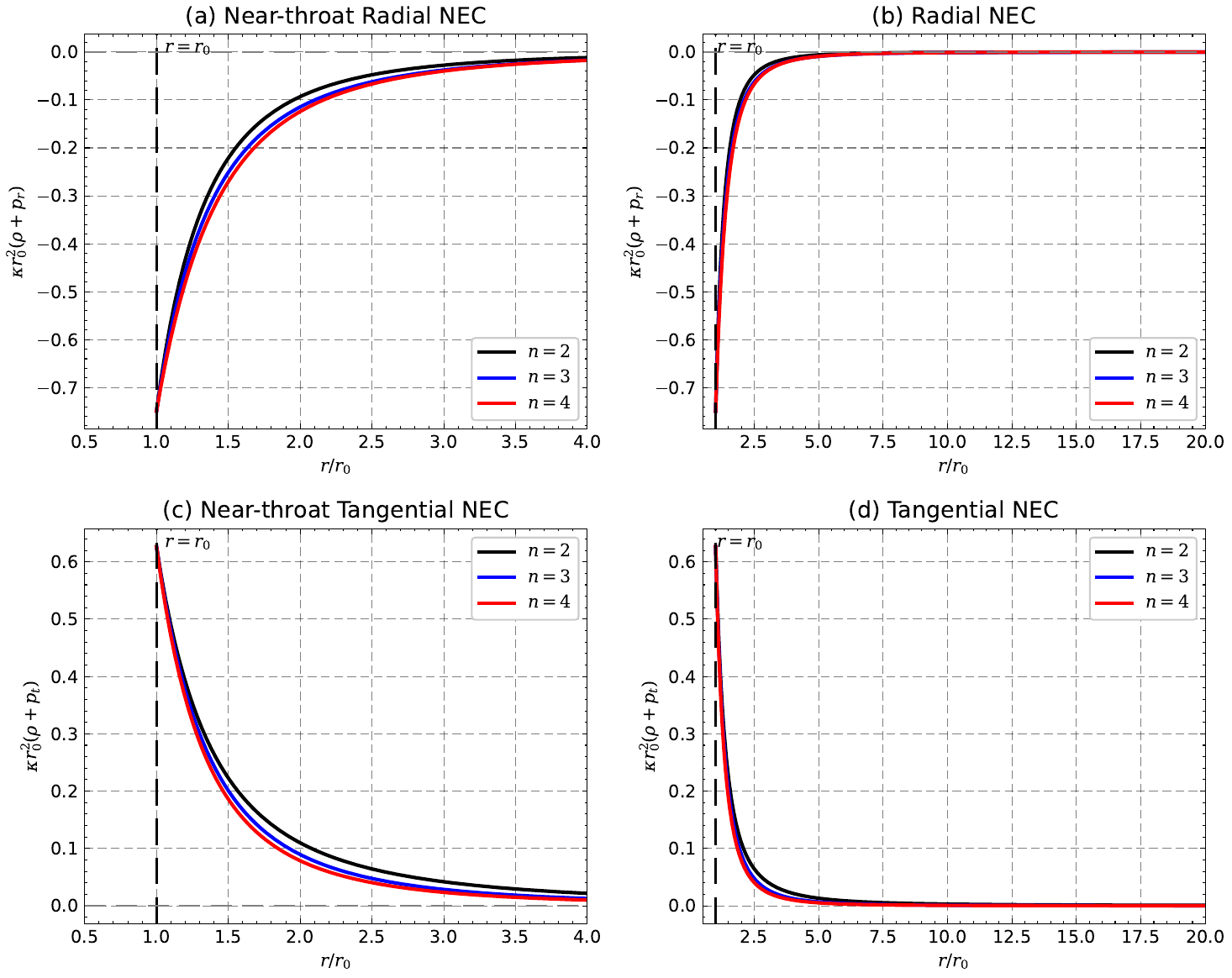}
    \caption{Radial and tangential null-energy conditions for the case of a logarithmic redshift function. The upper panels show the dimensionless radial NEC combination, $\kappa r_0^{2}\left(\rho+p_r\right)$, with the left panel providing a magnified view near the throat and the right panel displaying the behavior up to $r=20r_0$. The lower panels present the corresponding dimensionless tangential NEC combination, $\kappa r_0^{2}\left(\rho+p_t\right)$, with the same radial ranges. The curvature parameter is fixed at $\lambda=0.5$ and $v_0=0.05$, while tree values of the prescribed curvature exponent are considered, namely $n=2$, $n=3$ and $n=4$, illustrating the influence of the curvature profile on the energy conditions.}
    \label{fig:minha-imagem1}
\end{figure}

\begin{figure}[h!]
  \centering
  \begin{subfigure}[b]{0.49\textwidth}
    \includegraphics[width=\linewidth]{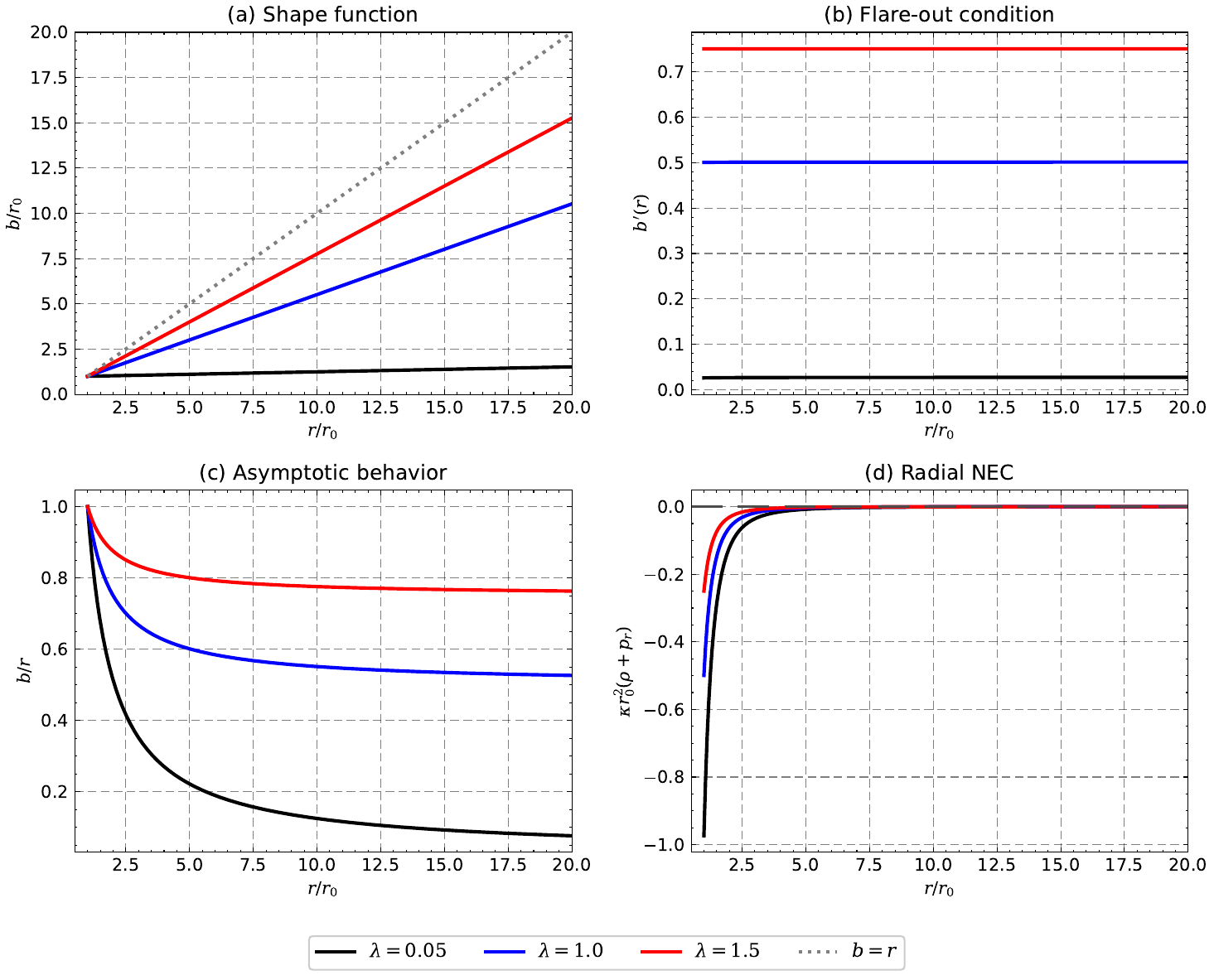}
    \label{fig:figura1}
  \end{subfigure}
   \hfill
  \begin{subfigure}[b]{0.49\textwidth}
    \includegraphics[width=\linewidth]{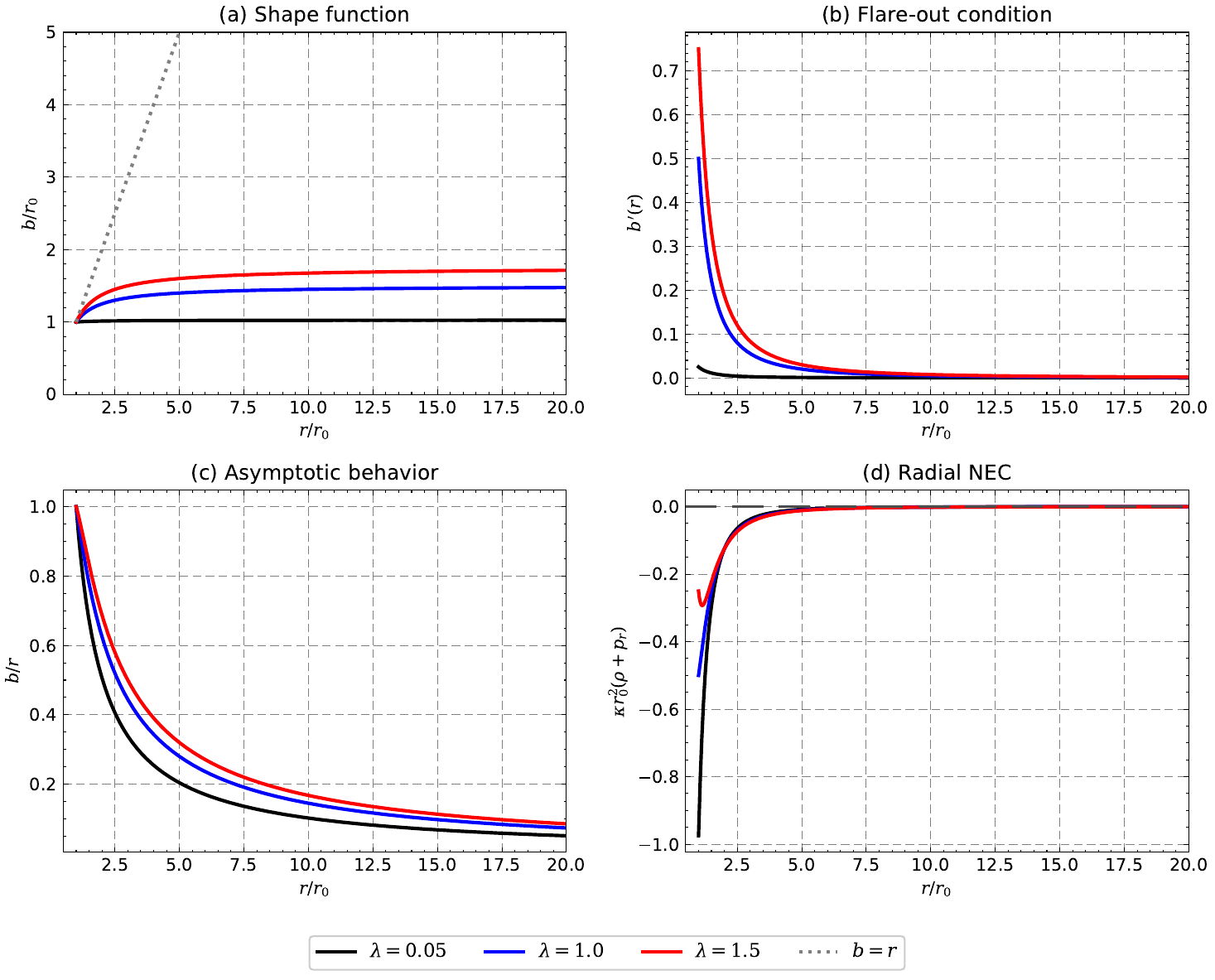}
    \label{fig:figura2}
  \end{subfigure}
  \caption{Geometrical and physical properties of the reconstructed wormhole supported by logarithmic redshift function for fixed $n=2$ (Left Panel), $n=4$ (Right Panel) and different values of the dimensionless curvature parameter $\lambda$. Panels (a)–(c) illustrate the influence of the curvature amplitude on the wormhole geometry, while panel (d) shows the corresponding behavior of the radial null energy condition. Increasing $\lambda$ produces a wider geometry and progressively reduces the amount of exotic matter required near the throat.}
  \label{fig:duas_figuras2}
\end{figure}

The numerical behavior of the reconstructed solutions is illustrated in Figs.~(\ref{fig:minha-imagem})--(\ref{fig:duas_figuras2}). We first analyze the influence of the curvature exponent $n$, followed by the dependence on the dimensionless curvature parameter $\lambda$.

\textbf{Figure}~\ref{fig:minha-imagem} presents the geometrical properties of the reconstructed wormhole supported by the logarithmic redshift function for different values of the curvature exponent $n$. As in the previous case, the throat condition $b(r_0)=r_0$ is automatically satisfied by construction, while the flare-out condition remains fulfilled throughout the parameter space considered. The different panels illustrate how the curvature exponent modifies the reconstructed geometry in the presence of a non-constant redshift function.
\begin{itemize}
\item The shape function $b(r)/r_0$, displayed in panel~\ref{fig:minha-imagem}(a), exhibits a monotonic increase from the throat while preserving the required wormhole geometry. As in the constant-redshift case, increasing the curvature exponent $n$ produces a steeper growth immediately outside the throat, reflecting the stronger localization of the curvature profile around the throat, as can also be seen in the magnified near-throat region shown in panel~\ref{fig:minha-imagem}(b). The overall behavior remains qualitatively unchanged with respect to the constant-redshift solution, indicating that the logarithmic redshift function introduces only small quantitative corrections to the reconstructed geometry for the adopted value of $v_0$. This behavior reflects the fact that the redshift contribution enters through terms proportional to $v_0^{2}$, which act as perturbative corrections to the geometry.
\item The asymptotic behavior, illustrated in panel (c), shows that the qualitative distinction between the reconstructed solutions remains unchanged in the presence of the logarithmic redshift function. For $n=2$, the ratio $b(r)/r$ approaches a non-vanishing constant at large distances, indicating that the reconstructed spacetime is not asymptotically flat. This behavior follows directly from Eq.~(\ref{eq:bxadmi}), where the curvature contribution becomes independent of the radial coordinate for $n=2$. In contrast, for larger values of the curvature exponent, represented here by $n=3$ and $n=4$, the curvature contribution decays rapidly with distance and the ratio $b(r)/r$ approaches the small constant $\xi/(\gamma+1)$, which is of order $v_0^2$.  Therefore, for the adopted value of $v_0$, the deviation from asymptotic flatness becomes extremely small. Moreover, in the limit $v_0\rightarrow0$, this residual contribution vanishes, consistently recovering the asymptotically flat behavior obtained for the constant-redshift solution.
\item Finally, panel (d) presents a magnified view of the derivative of the shape function in the vicinity of the throat. As expected, all reconstructed solutions satisfy the flare-out condition, with $b'(r_0)<1$. The different behaviors follow directly from Eq.~(\ref{eq:44.7}). For $n=2$, the curvature contribution remains constant, so that the radial dependence is governed only by the last term, whose effect is strongly suppressed by the small value of $\gamma$. Consequently, the derivative remains nearly constant throughout the displayed region. For $n=3$, the curvature contribution vanishes identically due to the factor $(3-n)$, leaving the derivative controlled by the constant term together with the small correction proportional to $r^{-(\gamma+1)}$. Finally, for $n=4$, the curvature contribution decays as $r^{-2}$, causing the derivative to decrease more rapidly away from the throat before approaching the residual constant $\xi/(\gamma+1)$. Therefore, although the logarithmic redshift function introduces small corrections to the flare-out behavior, the qualitative evolution of $b'(r)$ remains primarily controlled by the curvature exponent.
\end{itemize}

\textbf{Figure}~\ref{fig:minha-imagem1} illustrates the behavior of the null energy conditions for the reconstructed wormhole solutions for the case of logarithmic redshift function. 
\begin{itemize}
\item Panels (a) and (b) display the radial NEC in the vicinity of the throat and throughout the entire integration domain. As in the constant-redshift solution, the radial NEC remains violated over the whole spacetime, confirming that exotic matter is still required to sustain the wormhole geometry. However, the presence of the logarithmic redshift function modifies the analytical structure of Eq.~(\ref{eq:46.7}) through the additional contributions proportional to $v_0^2$. In particular, unlike the constant-redshift case, the curvature contribution no longer disappears for $n=2$, since its coefficient becomes $(2-n-2v_0^2)$ rather than simply $(2-n)$. Consequently, the radial NEC receives simultaneous contributions from the logarithmic redshift ($r^{-2}$), curvature ($r^{-n}$) and geometrical ($r^{-(\gamma+3)}$) sectors for all values of the curvature exponent, as previously discussed in subsection~(\ref{subsec:2}). Since the curvature scalar satisfies $R(r_0)=R_0$ independently of $n$, the exponent does not modify the curvature intensity at the throat, but rather the radial extent over which the curvature contributes to the geometry. Therefore, increasing $n$ causes the curvature contribution to decay more rapidly, weakening its compensating effect on the negative geometrical sector and leading to a stronger violation of the radial NEC away from the throat.
\item And, panels (c) and (d) display the tangential NEC in the vicinity of the throat and throughout the entire integration domain. In contrast to the radial sector, no violation of the tangential NEC is observed for any of the reconstructed solutions. This behavior follows directly from Eq.~(\ref{eq:47.7}), where the tangential NEC receives simultaneous contributions from the logarithmic redshift $r^{-2}$, curvature $r^{-n}$ and geometrical terms $r^{-(\gamma+3)}$ for all values of the curvature exponent. Unlike the radial case, the curvature contribution never vanishes identically, remaining present even for $n=2$. For larger values of $n$, this contribution decays more rapidly with the radial coordinate, while the remaining terms dominate the asymptotic behavior.
\end{itemize}

\textbf{Figure}~\ref{fig:duas_figuras2} illustrates the influence of the dimensionless curvature parameter $\lambda$ on the reconstructed wormhole geometry supported by the logarithmic redshift function for fixed values of the curvature exponent, namely $n=2$ (left panels) and $n=4$ (right panels). Since $\lambda$ determines the amplitude of the curvature profile, varying this parameter allows one to investigate how the geometry and the corresponding energy conditions respond to changes in the curvature scale while preserving the same halo redshift function.
\begin{itemize}
\item Panels (a) show the behavior of the reconstructed shape function for different values of the dimensionless curvature parameter. As in the constant-redshift configuration, increasing $\lambda$ systematically enlarges the reconstructed wormhole geometry while preserving the throat condition $b(r_0)=r_0$. The influence of $\lambda$ is observed as a steeper growth of the shape function immediately outside the throat, reflecting the stronger contribution of the curvature sector to the reconstructed geometry. This qualitative behavior is preserved for both $n=2$ and $n=4$, indicating that the logarithmic redshift function introduces only perturbative corrections, without modifying the geometrical role played by the dimensionless curvature parameter.
\item Panels (b) illustrate the behavior of the derivative of the shape function for different values of the dimensionless curvature parameter. As expected from Eq.~(\ref{eq:44.7}), increasing $\lambda$ systematically raises the value of $b'(r)$ throughout the radial domain, reflecting the stronger contribution of the curvature sector to the flare-out behavior. Nevertheless, two qualitatively distinct regimes remain evident. For $n=2$, the curvature contribution is constant and the radial dependence is governed only by the small correction proportional to $r^{-(\gamma+1)}$, causing the derivative to remain nearly constant over the displayed interval. In contrast, for $n=4$, the curvature contribution decreases as $r^{-2}$, leading to a rapid reduction of $b'(r)$ away from the throat before approaching the residual constant $\xi/(\gamma+1)$. Therefore, while $\lambda$ controls the overall magnitude of the flare-out condition, the radial evolution of the derivative continues to be determined primarily by the curvature exponent.
\item Panels (c) illustrate the asymptotic behavior of the reconstructed solutions through the ratio $b(r)/r$. As in the constant-redshift configuration, the dimensionless curvature parameter modifies the amplitude of the reconstructed geometry without changing its asymptotic class. For $n=2$, increasing $\lambda$ raises the non-vanishing asymptotic value of $b(r)/r$, indicating that the reconstructed spacetime remains non-asymptotically flat. In contrast, for $n=4$, the curvature contribution decays as $r^{-2}$ and the ratio $b(r)/r$ approaches the small residual constant $\xi/(\gamma+1)$, independently of the particular value of $\lambda$. Consequently, the logarithmic redshift function preserves the qualitative asymptotic behavior of the reconstructed solutions, while introducing only perturbative corrections of order $v_0^2$ with respect to the constant-redshift case.
\item Finally, panels (d) present the behavior of the radial null energy condition for different values of the dimensionless curvature parameter. For both values of the curvature exponent, increasing $\lambda$ systematically reduces the magnitude of the NEC violation, causing the curves to approach the null value while remaining negative throughout the entire integration domain. This behavior follows directly from Eq.~(\ref{eq:46.7}), where the curvature contribution grows proportionally with $\lambda$, partially compensating the negative geometrical contribution responsible for sustaining the wormhole. Consequently, larger values of the dimensionless curvature parameter reduce the amount of exotic matter required to support the reconstructed geometry. Nevertheless, this improvement remains naturally bounded by the flare-out condition, which restricts the admissible values of $\lambda$. Therefore, within the logarithmic redshift function, the dimensionless curvature parameter plays a dual role: it controls both the geometrical deformation of the wormhole and the amount of exotic matter necessary to sustain the solution, consistently with the behavior previously observed for the constant-redshift configuration.
\end{itemize}

Overall, the numerical analysis shows that the qualitative role of the parameters remains unchanged in the presence of the logarithmic redshift function. The curvature exponent $n$ determines the asymptotic class and radial evolution of the reconstructed geometry, whereas the dimensionless curvature parameter $\lambda$ controls the intensity of both the geometrical deformation and the violation of the radial null energy condition. The logarithmic redshift function introduces only perturbative corrections proportional to $v_0^2$, preserving the overall physical properties of the reconstructed wormhole solutions.

\subsection{Case of Inverse-Radial Redshift Function, $\Phi\left(r\right)=r_0/r$.}
In contrast to the previous redshift configurations, the inverse-radial profile does not lead to simple closed-form expressions for the effective matter sector and the asymptotic behavior of the reconstructed geometry. Although the local properties at the throat can be determined analytically (see Eqs.~(\ref{eq:bprime_throat_exp}), (\ref{eq:nec_exp}) and (\ref{eq:nect_exp})), the global solution is naturally expressed in terms of the integral operators $J_k(\epsilon)$, making the complete analytical discussion cumbersome. Consequently, the physical properties of the reconstructed wormhole are more transparently investigated through a numerical analysis of the geometry and the corresponding null energy conditions.
\begin{figure}[h!]
    \centering
    \includegraphics[width=0.8\textwidth]{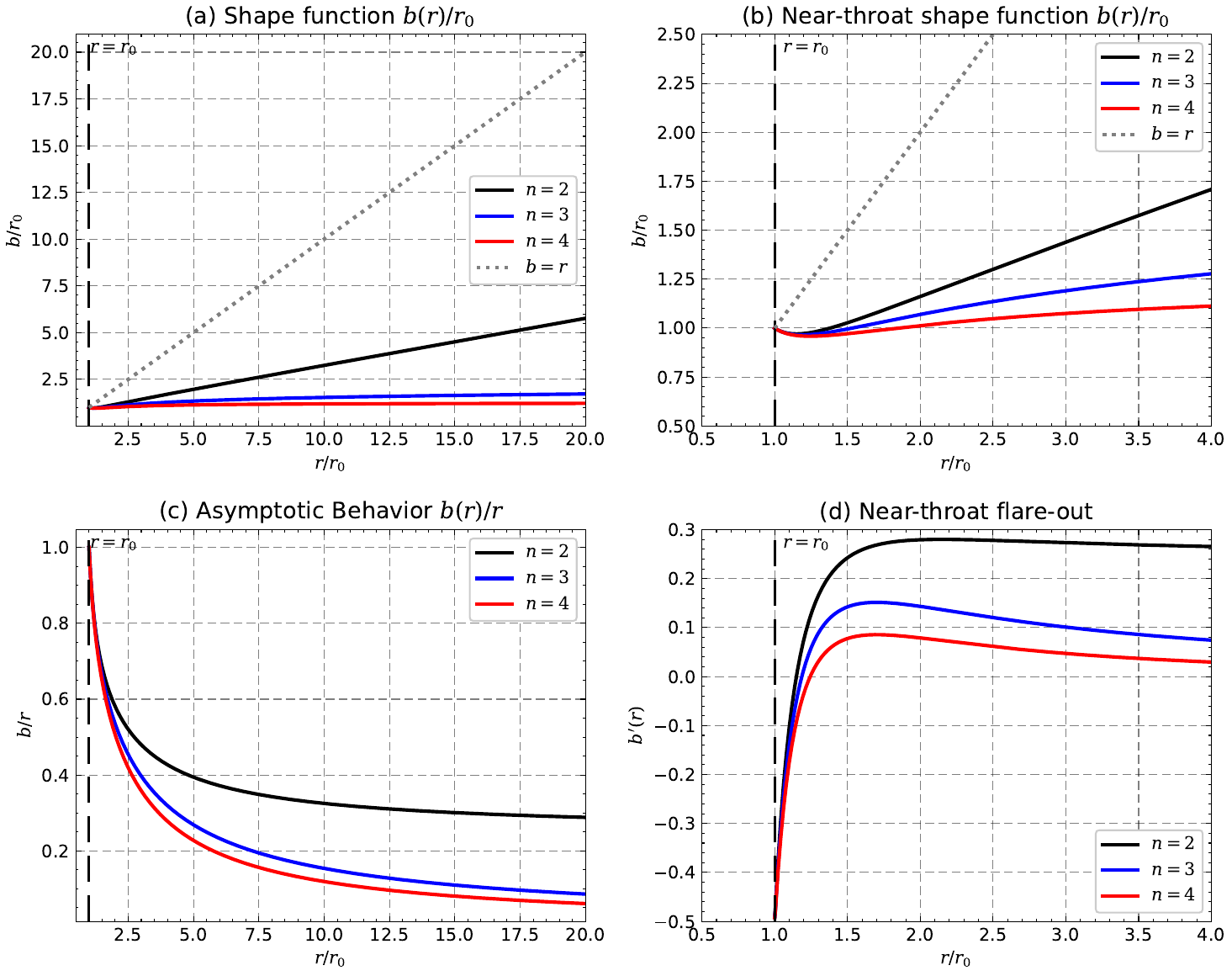}
    \caption{Geometric properties of the reconstructed wormhole supported by redshift function $\Phi(r)=r_0/r$ for $\lambda=0.5$ and different values of $n$. Panel (a) shows the shape function, panel (b) illustrates near-throat shape function, while panel (c) its asymptotic behavior through the ratio $b(r)/r$, (d) illustrates near-throat flare-out condition.}
    \label{fig:minha-imagem10}
\end{figure}

As in the previous sections, we first analyze the influence of the curvature exponent $n$ on the reconstructed geometry and NEC, followed by the role of the dimensionless curvature parameter $\lambda$.

\textbf{Figure}~(\ref{fig:minha-imagem10}) presents the geometrical properties of the reconstructed wormhole supported by the inverse-radial redshift function for different values of the curvature exponent. 

\begin{itemize}
\item Unlike the previous redshift configurations, the inverse-radial redshift function introduces a qualitatively new feature near the throat. As predicted analytically by Eq.~(\ref{eq:bprime_throat_exp}), the negative value of $b'(r_0)$ for $\lambda<1$ causes the shape function to decrease immediately outside the throat, as shown in panel (b). This local minimum is absent in the previous reconstructions. Away from the throat, however, the influence of the curvature exponent remains qualitatively unchanged, with larger values of $n$ producing a steeper growth of the reconstructed geometry.
\item Panel (c) shows that the inverse-radial redshift function does not modify the asymptotic classification of the reconstructed solutions. As in the previous redshift configurations, the case $n=2$ remains non-asymptotically flat, since $b(r)/r$ approaches a non-vanishing constant at large distances. Conversely, for $n\ge3$ the ratio decreases monotonically towards zero, indicating that the reconstructed geometry becomes asymptotically flat. Therefore, the inverse-radial redshift function changes the local behavior of the wormhole near the throat without altering its global asymptotic structure.
\item Finally, panel (d) illustrates the radial derivative of the shape function in the vicinity of the throat. Unlike the previous redshift configurations, the flare-out function is no longer necessarily positive at the throat. Instead, as predicted analytically by Eq.~(\ref{eq:bprime_throat_exp}), its value is completely determined by the dimensionless curvature parameter through $b'(r_0)=\lambda-1$. For the adopted value $\lambda=0.5$, the derivative is negative at the throat, explaining the initial decrease of the shape function observed in panel (b). As the radial coordinate increases, however, the curvature contribution becomes progressively subdominant and the derivative changes sign, allowing the shape function to recover its monotonic growth. Increasing the curvature exponent accelerates the decay of the curvature sector, causing the transition to positive values of $b'(r)$ to occur closer to the throat.
\end{itemize}
\begin{figure}[t!]
\centering
    \includegraphics[width=0.8\textwidth]{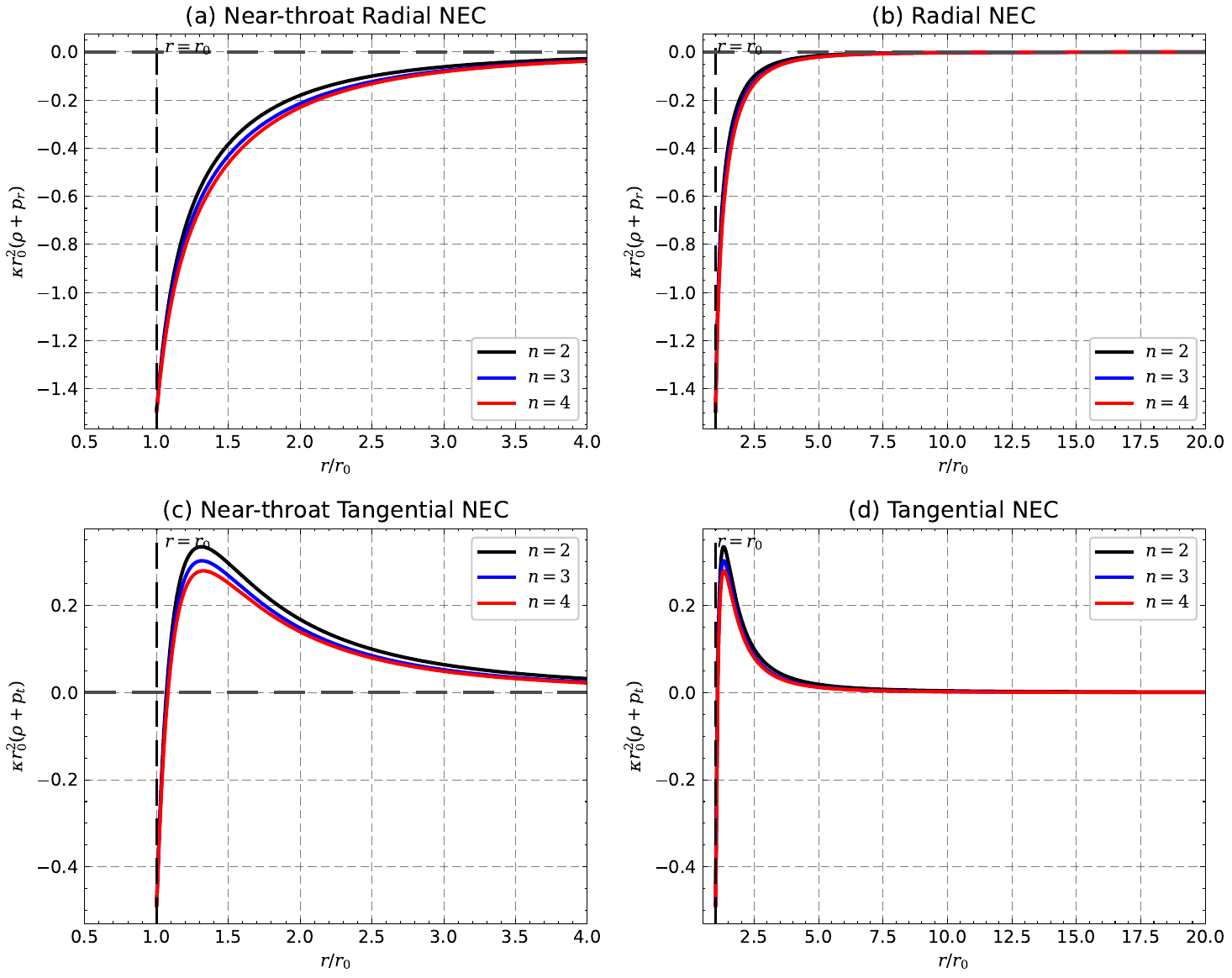}
    \caption{Radial and tangential null-energy conditions for the case of a inverse-radial redshift function. The upper panels show the dimensionless radial NEC combination, $\kappa r_0^{2}\left(\rho+p_r\right)$, with the left panel providing a magnified view near the throat and the right panel displaying the behavior up to $r=20r_0$. The lower panels present the corresponding dimensionless tangential NEC combination, $\kappa r_0^{2}\left(\rho+p_t\right)$, with the same radial ranges. The curvature parameter is fixed at $\lambda=0.5$, while tree values of the prescribed curvature exponent are considered, namely $n=2$, $n=3$ and $n=4$, illustrating the influence of the curvature profile on the energy conditions.}
    \label{fig:minha-imagem11}
\end{figure}

\textbf{Figure}~\ref{fig:minha-imagem11} presents the behavior of the radial and tangential NEC for the inverse-radial redshift reconstruction. As in the previous configurations, the numerical analysis allows us to investigate the spatial distribution of the effective exotic matter supporting the wormhole geometry.
\begin{itemize}
\item Panels (a) and (b) show the radial null energy condition near the throat and throughout the integration domain. The radial NEC remains negative everywhere, confirming that exotic matter is still required to sustain the reconstructed wormhole. As the curvature exponent increases, the curvature contribution becomes progressively more localized around the throat. Consequently, larger values of $n$ produce more negative radial NEC profiles, since the curvature contribution decays more rapidly and becomes unable to compensate the geometrical sector at larger distances.
\item Unlike the previous redshift configurations, panels (c) and (d) reveal a qualitatively different behavior for the tangential NEC. A narrow region of NEC violation appears immediately outside the throat, after which $\kappa\left(\rho+p_t\right)$ rapidly changes sign and remains positive throughout the remaining integration domain. This localized violation follows directly from the inverse-radial redshift function, which modifies the near-throat geometry through the negative value of $b'(r_0)$ and changes the balance between the curvature and geometrical sectors in Eq.~(\ref{eq:nect_exp}). Nevertheless, this violation remains highly localized around the throat, while the tangential NEC rapidly recovers positive values outside this region. This indicates that the inverse-radial redshift modifies the effective matter distribution only locally, without affecting the global behavior of the reconstructed solution.
\end{itemize}
\begin{figure}[h!]
  \centering
  \begin{subfigure}[b]{0.49\textwidth}
    \includegraphics[width=\linewidth]{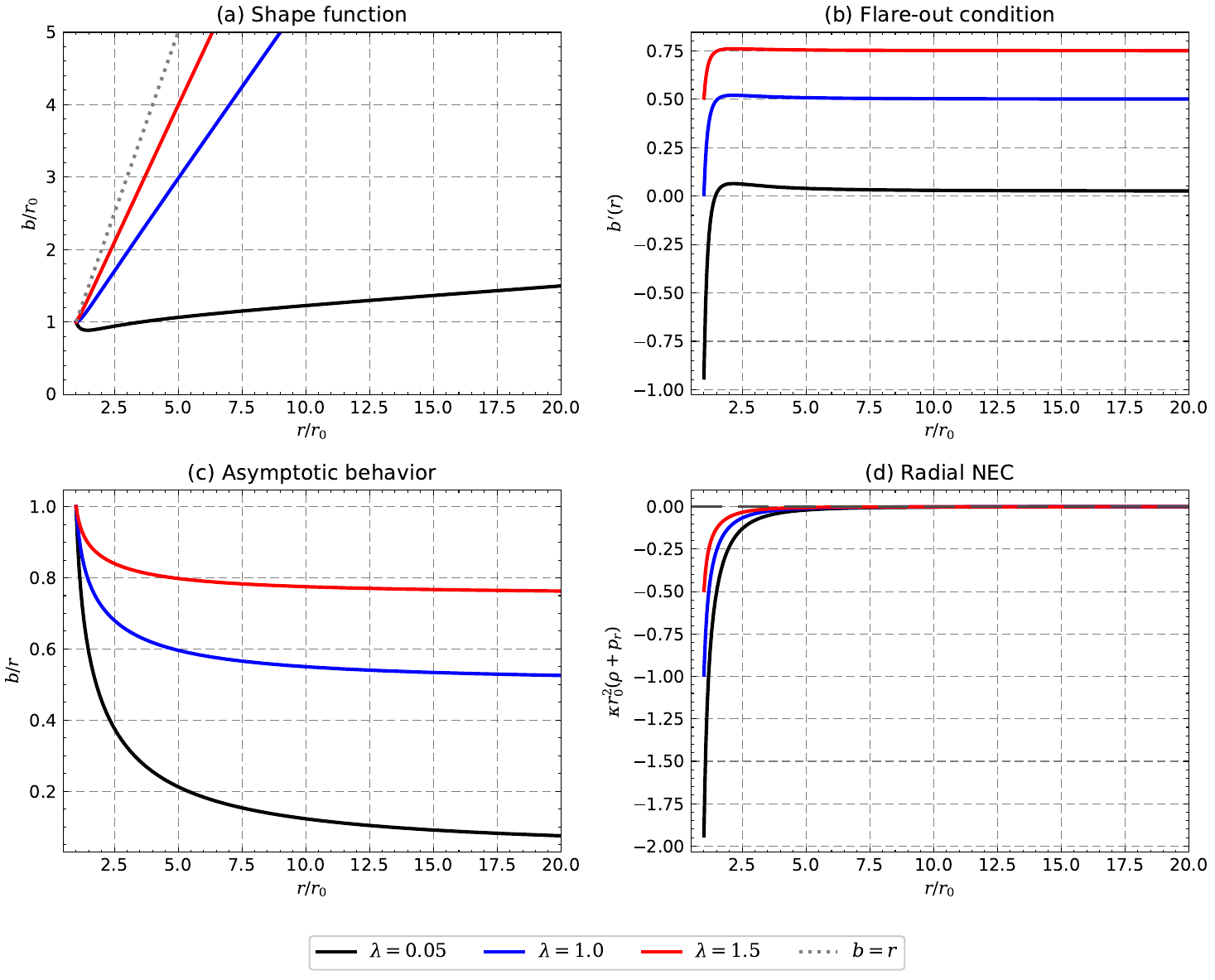}
    \label{fig:figura1}
  \end{subfigure}
   \hfill
  \begin{subfigure}[b]{0.49\textwidth}
    \includegraphics[width=\linewidth]{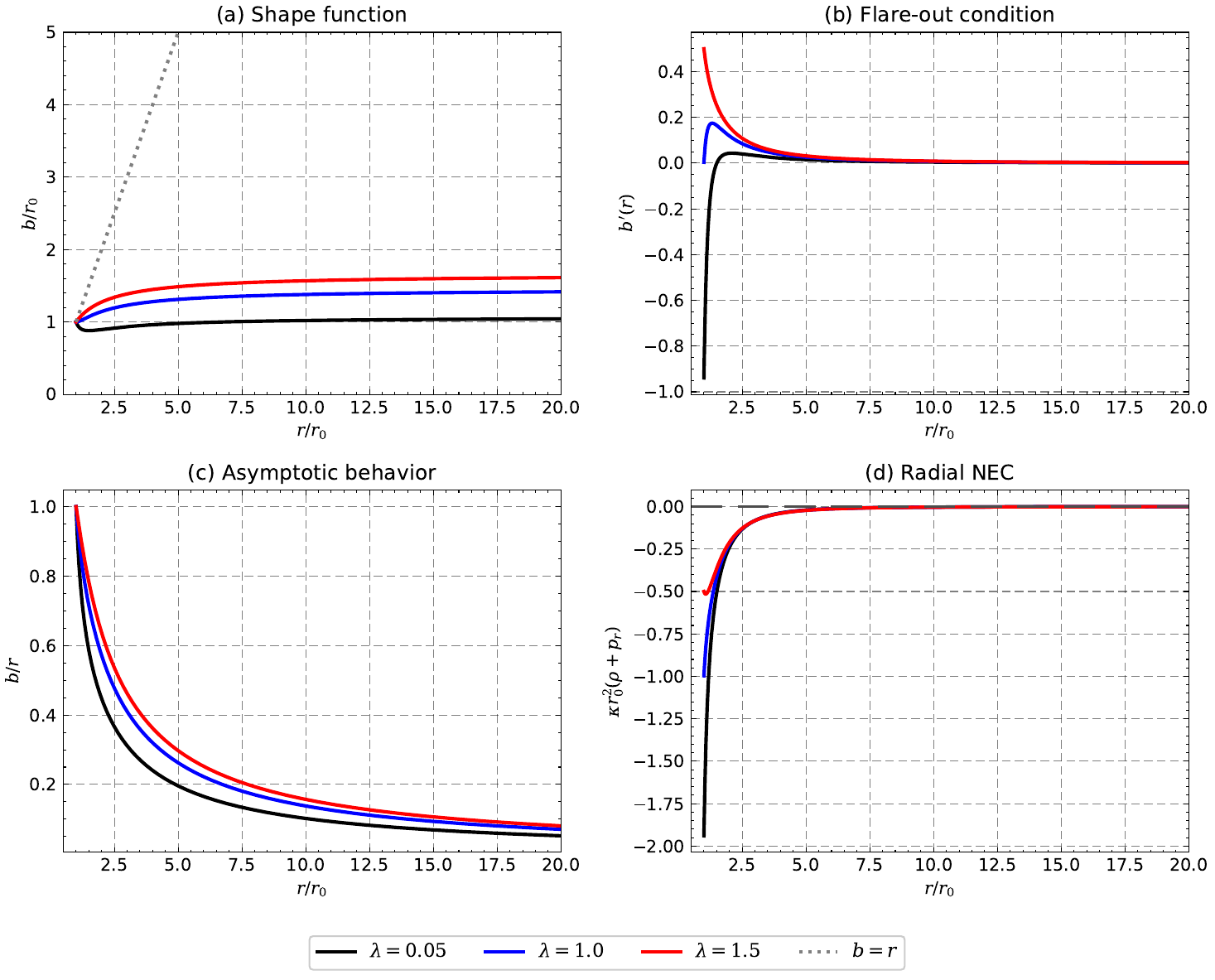}
    \label{fig:figura2}
  \end{subfigure}
  \caption{Geometrical and physical properties of the reconstructed wormhole supported by $\Phi(r)=r_0/r$ redshift function for fixed $n=2$ (Left Panel), $n=4$ (Right Panel) and different values of the dimensionless curvature parameter $\lambda$. Panels (a)–(c) illustrate the influence of the curvature amplitude on the wormhole geometry, while panel (d) shows the corresponding behavior of the radial null energy condition. Increasing $\lambda$ produces a wider geometry and progressively reduces the amount of exotic matter required near the throat.}
  \label{fig:duas_figuras12}
\end{figure}

\textbf{Figure}~\ref{fig:duas_figuras12} illustrates the influence of the dimensionless curvature parameter on the reconstructed geometry and the radial NEC for the inverse-radial redshift function. As in the previous reconstructions, the analysis is performed for two representative values of the curvature exponent, allowing us to distinguish the effects associated with the curvature amplitude from those related to its radial decay.
\begin{itemize}
\item Panels (a) and (b) illustrate the influence of the dimensionless curvature parameter on the reconstructed geometry. Increasing $\lambda$ systematically enlarges the shape function and raises the derivative $b'(r)$ throughout the radial domain. In particular, since the slope at the throat is analytically given by $b'(r_0)=\lambda-1$, larger values of the curvature amplitude progressively suppress the initial decrease of the shape function observed in the inverse-radial reconstruction. For sufficiently large values of $\lambda$, the derivative becomes positive already at the throat, restoring the monotonic growth of the geometry immediately outside the throat. This transition occurs for both values of the curvature exponent, although the subsequent radial evolution remains controlled by $n$: the $n=2$ solutions preserve the influence of the curvature sector over larger distances, whereas for $n=4$ the curvature effects become rapidly localized around the throat.
\item Panels (c) shows that varying the dimensionless curvature parameter does not modify the asymptotic behavior of the reconstructed solutions. As in the previous redshift configurations, the asymptotic class remains entirely determined by the curvature exponent. Increasing $\lambda$ changes only the amplitude of the shape function, whereas the distinction between the non-asymptotically flat solution ($n=2$) and the effectively asymptotically flat solution ($n=4$) is completely preserved.
\item Panels (d) show the behavior of the radial NEC for different values of the dimensionless curvature parameter. As discussed in previous cases, when we assume a redshift function of the inverse-radial type, for both values of the curvature exponent, increasing $\lambda$ systematically reduces the magnitude of the NEC violation, causing the curves to approach the null value while remaining negative throughout the radial domain. Consequently, larger values of the curvature parameter decrease the amount of exotic matter required to sustain the wormhole geometry.
\end{itemize}

Overall, the inverse-radial redshift reconstruction preserves the fundamental roles played by the reconstruction parameters while introducing a distinctive local geometrical feature. The curvature exponent $n$ continues to control the radial extension of the curvature sector and, consequently, the asymptotic behavior of the reconstructed spacetime. In contrast, the dimensionless curvature parameter $\lambda$ governs the intensity of the curvature contribution, regulating both the local geometry near the throat and the amount of exotic matter required to sustain the wormhole. Unlike the previous redshift configurations, however, the inverse-radial profile exhibits a transition between an initially decreasing and a monotonically increasing shape function, entirely controlled by the curvature amplitude through the analytical relation $b'(r_0)=\lambda-1$.

\section{Conclusion}\label{sec:6}

In this work, we have developed a curvature reconstruction framework for static and spherically symmetric traversable wormholes within non-conservative unimodular gravity. Exploiting the fact that $\mathrm{NUG}$ does not impose an algebraic relation between the Ricci scalar and the trace of the energy-momentum tensor, we treated the Ricci scalar as an independently prescribed geometric quantity and reconstructed the wormhole geometry directly from a prescribed curvature profile. This approach provides an alternative reconstruction strategy to the conventional procedure adopted in the literature, where either the matter distribution or the metric functions are specified a priori.

The reconstruction framework was applied to a common power-law curvature profile combined with three distinct redshift functions, leading to different classes of traversable wormhole geometries. Although all reconstructed solutions satisfy the standard flare-out condition and necessarily violate the radial null energy condition at the throat, in agreement with previous results obtained in unimodular gravity, they exhibit distinct geometrical properties, asymptotic behaviors and energy-condition profiles.

Beyond the reconstruction procedure itself, the present analysis reveals two noteworthy features. First, although the radial null energy condition remains necessarily violated at the throat, in agreement with previous analyses based on trace-free gravitational field equations \citep{Cataldo_2026, Pastén_2026}, the dimensionless curvature parameter $\lambda$ directly controls the magnitude of this violation. As $\lambda$ approaches its critical value, the exotic matter required to support the wormhole becomes progressively weaker, indicating that the curvature sector actively contributes to the physical viability of the reconstructed solutions, as demonstrated throughout our numerical analysis. 

A second interesting consequence concerns the geometrical flare-out condition. Although the reconstructed wormholes are generated from different values of the power-law exponent $n$, the throat condition depends exclusively on the dimensionless curvature parameter $\lambda$, remaining insensitive to the exponent that governs the asymptotic behavior of the prescribed curvature profile, see Eqs.~(\ref{eq:flare}), (\ref{eq:45.7}) and (\ref{eq:flare_exp}). Whether this feature is a particular consequence of the adopted power-law curvature ansatz or represents a more general property of curvature-based reconstruction constitutes an interesting open question.

Another noteworthy result concerns the power-law exponent $n$ naturally separates different asymptotic regimes, with the critical configuration $n=2$ emerging as the transition between asymptotically flat wormholes and geometries exhibiting residual asymptotic deformations. This behavior follows directly from the asymptotic structure of the reconstructed shape function and suggests that the prescribed curvature profile governs not only the local geometry near the throat but also the global structure of the spacetime.

Although the logarithmic redshift profile produces only perturbative corrections to the reconstructed geometry, its influence on the geodesic structure is expected to be considerably more significant, since the redshift function enters explicitly into the effective potential governing particle and photon motion. A detailed investigation of the corresponding geodesic dynamics is currently in progress and will be presented in a future publication.

The present work also opens several directions for future investigation. First, the dynamical stability of the reconstructed wormhole solutions remains to be analyzed, since the present study was restricted to their geometrical reconstruction and energy conditions. A stability analysis under linear perturbations will be essential for assessing the physical viability of these configurations. Second, although the non-conservation equation does not explicitly participate in the reconstruction procedure, it provides the theoretical consistency that keeps the algebraic relation $R$ and $T$ is open. This suggests that its physical role may become considerably more relevant from a thermodynamical perspective, where the non-conservative character of the theory could be directly associated with irreversible energy exchange between matter and geometry. Finally, the reconstruction scheme developed here was implemented for a power-law curvature profile. Extending this approach to more general curvature prescriptions may clarify whether the geometrical properties identified in this work, particularly the dependence of the flare-out condition on the dimensionless curvature parameter and the separation between local and asymptotic geometrical behavior, represent specific features of the adopted ansatz or constitute generic properties of curvature-based reconstruction in non-conservative unimodular gravity. A particularly interesting direction concerns the constant-curvature case ($n=0$), which naturally leads to de Sitter or anti-de Sitter wormhole geometries and therefore belongs to a qualitatively different class of solutions from those investigated here.

We hope that the curvature reconstruction framework developed in this work may provide a useful starting point for investigating compact objects from prescribed Ricci scalar profiles, opening new possibilities for geometrical reconstruction in non-conservative unimodular gravity.
\begin{acknowledgments}
We thank CNPq, FAPES and CAPES for partial financial support.
\end{acknowledgments}
 \bibliography{bibli1}{}

@article{universe9120515,
AUTHOR = {Fabris, Júlio César and Daouda, Mahamadou Hamani and Velten, Hermano},
TITLE = {Spherically Symmetric Configurations in Unimodular Gravity},
JOURNAL = {Universe},
VOLUME = {9},
YEAR = {2023},
NUMBER = {12},
ARTICLE-NUMBER = {515},
URL = {https://www.mdpi.com/2218-1997/9/12/515},
ISSN = {2218-1997},
DOI = {10.3390/universe9120515}
}

@article{Anderson:1971pn,
    author = "Anderson, J. L. and Finkelstein, D.",
    title = "{Cosmological constant and fundamental length}",
    doi = "10.1119/1.1986321",
    journal = "Am. J. Phys.",
    volume = "39",
    pages = "901--904",
    year = "1971"
}

@ARTICLE{1991JMP....32.1337N,
       author = {{Ng}, Y.~J. and {van Dam}, H.},
        title = "{Unimodular theory of gravity and the cosmological constant.}",
      journal = {Journal of Mathematical Physics},
         year = 1991,
        month = may,
       volume = {32},
       number = {5},
        pages = {1337-1340},
          doi = {10.1063/1.529283},
       adsurl = {https://ui.adsabs.harvard.edu/abs/1991JMP....32.1337N}
}

@article{RevModPhys.61.1,
  title = {The cosmological constant problem},
  author = {Weinberg, Steven},
  journal = {Rev. Mod. Phys.},
  volume = {61},
  issue = {1},
  pages = {1--23},
  numpages = {0},
  year = {1989},
  month = {Jan},
  publisher = {American Physical Society},
  doi = {10.1103/RevModPhys.61.1},
}

@article{Ellis:2013uxa,
    author = "Ellis, George F R",
    title = "{The Trace-Free Einstein Equations and inflation}",
    eprint = "1306.3021",
    archivePrefix = "arXiv",
    primaryClass = "gr-qc",
    doi = "10.1007/s10714-013-1619-5",
    journal = "Gen. Rel. Grav.",
    volume = "46",
    pages = "1619",
    year = "2014"
}

@article{Ellis:2010uc,
    author = "Ellis, George F. R. and van Elst, Henk and Murugan, Jeff and Uzan, Jean-Philippe",
    title = "{On the Trace-Free Einstein Equations as a Viable Alternative to General Relativity}",
    eprint = "1008.1196",
    archivePrefix = "arXiv",
    primaryClass = "gr-qc",
    doi = "10.1088/0264-9381/28/22/225007",
    journal = "Class. Quant. Grav.",
    volume = "28",
    pages = "225007",
    year = "2011"
}

@book{Weinberg,
   title =     {Gravitation and cosmology: principles and applications of the general theory of relativity},
   author =    {Weinberg, S},
   publisher = {Wiley},
   isbn =      {9780471925675,0-471-92567-5},
   year =      {1972},

}

@article{Alvarez:2023eqo,
    author = "{\'A}lvarez, Enrique and Anero, Jes{\'u}s and S{\'a}nchez-Ruiz, Irene",
    title = "{The origin of the cosmological constant in unimodular gravity}",
    eprint = "2310.16522",
    archivePrefix = "arXiv",
    primaryClass = "hep-th",
    reportNumber = "IFT-UAM/CSIC-23-133, FTUAM-23-xx",
    doi = "10.1140/epjc/s10052-024-12651-7",
    journal = "Eur. Phys. J. C",
    volume = "84",
    number = "3",
    pages = "287",
    year = "2024"
}

@book{Carroll:2004st,
    author = "Carroll, Sean M.",
    title = "{Spacetime and Geometry}: {An Introduction to General Relativity}",
    doi = "10.1017/9781108770385",
    isbn = "978-0-8053-8732-2, 978-1-108-48839-6, 978-1-108-77555-7",
    publisher = "Cambridge University Press",
    month = "7",
    year = "2019"
}

@article{Cataldo_2026,
   title={Can Wormhole Spacetimes in Unimodular Gravity Be Supported by Ordinary Matter? A General Proof of the Exotic Matter Requirement},
   volume={15},
   ISSN={2075-1680},
   url={http://dx.doi.org/10.3390/axioms15040244},
   DOI={10.3390/axioms15040244},
   number={4},
   journal={Axioms},
   publisher={MDPI AG},
   author={Cataldo, Mauricio and Cruz, Norman and Salgado, Patricio},
   year={2026},
   month=mar, pages={244} }

@book{91143000,
   title =     {Theoretical astrophysics. Galaxies and cosmology Volume 3},
   author =    {T. Padmanabhan},
   publisher = {Cambridge University Press},
   isbn =      {0521566304; 9780521566308},
   year =      {2002},
   edition =   {First},
   }

@article{Begeman:1991iy,
    author = "Begeman, K. G. and Broeils, A. H. and Sanders, R. H.",
    title = "{Extended rotation curves of spiral galaxies: Dark haloes and modified dynamics}",
    doi = "10.1093/mnras/249.3.523",
    journal = "Mon. Not. Roy. Astron. Soc.",
    volume = "249",
    pages = "523",
    year = "1991"
}

@book{bookBinney,
   title =     {Galactic Dynamics},
   author =    {James Binney, Scott Tremaine},
   publisher = {Princeton University Press},
   isbn =      {0691130264; 9780691130262; 0691130272; 9780691130279},
   year =      {2008},
   series =    {Princeton Series in Astrophysics},
   edition =   {2nd}
  }

@book{Misner:1973prb,
    author = "Misner, Charles W. and Thorne, K. S. and Wheeler, J. A.",
    title = "{Gravitation}",
    isbn = "978-0-7167-0344-0, 978-0-691-17779-3",
    publisher = "W. H. Freeman",
    address = "San Francisco",
    year = "1973"
}

@article{Navarro_1997,
   title={A Universal Density Profile from Hierarchical Clustering},
   volume={490},
   ISSN={1538-4357},
   url={http://dx.doi.org/10.1086/304888},
   DOI={10.1086/304888},
   number={2},
   journal={The Astrophysical Journal},
   publisher={American Astronomical Society},
   author={Navarro, Julio F. and Frenk, Carlos S. and White, Simon D. M.},
   year={1997},
   month=Dec, pages={493–508} 
   
}

@book{Sparke_Gallagher_2007,
  title     = {Galaxies in the Universe: An Introduction},
  author    = {Sparke, Linda S. and Gallagher, John S.},
  year      = {2007},
  edition   = {2nd},
  publisher = {Cambridge University Press},
  address   = {Cambridge, UK},
  isbn      = {978-0521671866}
}

@article{Sofue_2001,
   title={Rotation Curves of Spiral Galaxies},
   volume={39},
   ISSN={1545-4282},
   url={http://dx.doi.org/10.1146/annurev.astro.39.1.137},
   DOI={10.1146/annurev.astro.39.1.137},
   number={1},
   journal={Annual Review of Astronomy and Astrophysics},
   publisher={Annual Reviews},
   author={Sofue, Yoshiaki and Rubin, Vera},
   year={2001},
   month="sep", pages={137–174} 
}

@article{Morris:1988cz,
    author = "Morris, M. S. and Thorne, K. S.",
    title = "{Wormholes in space-time and their use for interstellar travel: A tool for teaching general relativity}",
    doi = "10.1119/1.15620",
    journal = "Am. J. Phys.",
    volume = "56",
    pages = "395--412",
    year = "1988"
}

@article{PhysRev.48.73,
  title = {The Particle Problem in the General Theory of Relativity},
  author = {Einstein, A. and Rosen, N.},
  journal = {Phys. Rev.},
  volume = {48},
  issue = {1},
  pages = {73--77},
  numpages = {0},
  year = {1935},
  month = {Jul},
  publisher = {American Physical Society},
  doi = {10.1103/PhysRev.48.73},
  url = {https://link.aps.org/doi/10.1103/PhysRev.48.73}
}

@book{Visser:1995cc,
  title={Lorentzian Wormholes: From Einstein to Hawking},
  author={Visser, Matt},
  year={1995},
  publisher={American Institute of Physics},
  address={Woodbury, NY},
  isbn={9781563966538},
  series={AIP Series in Computational and Applied Mathematical Physics}
}

@book{Sagan1985,
  author    = {Carl Sagan},
  title     = {Contact},
  publisher = {Simon \& Schuster},
  year      = {1985},
  isbn      = { 978-0671434007}
}

@book{Thorne1994,
  author    = {Kip S. Thorne},
  title     = {Black Holes and Time Warps: Einstein's Outrageous Legacy},
  publisher = {W. W. Norton \& Company},
  year      = {1994},
  isbn      = {978-0393035056}
}

@article{PhysRevD.56.4745,
  title = {Geometric structure of the generic static traversable wormhole throat},
  author = {Hochberg, David and Visser, Matt},
  journal = {Phys. Rev. D},
  volume = {56},
  issue = {8},
  pages = {4745--4755},
  numpages = {0},
  year = {1997},
  month = {Oct},
  publisher = {American Physical Society},
  doi = {10.1103/PhysRevD.56.4745},
  url = {https://link.aps.org/doi/10.1103/PhysRevD.56.4745}
}

@article{Ellis:1973yv,
    author = "Ellis, H. G.",
    title = "{Ether flow through a drainhole - a particle model in general relativity}",
    doi = "10.1063/1.1666161",
    journal = "J. Math. Phys.",
    volume = "14",
    pages = "104--118",
    year = "1973"
}

@article{Bronnikov:1973fh,
    author = "Bronnikov, K. A.",
    title = "{Scalar-tensor theory and scalar charge}",
    journal = "Acta Phys. Polon. B",
    volume = "4",
    pages = "251--266",
    year = "1973"
}

@article{Lobo:2005us,
    author = "Lobo, Francisco S. N.",
    title = "{Phantom energy traversable wormholes}",
    journal = "Phys. Rev. D",
    volume = "71",
    pages = "084011",
    year = "2005",
    eprint = "gr-qc/0502099",
    archivePrefix = "arXiv",
    doi = "10.1103/PhysRevD.71.084011",
    url = "https://link.aps.org/doi/10.1103/PhysRevD.71.084011"
}

@article{PhysRevD.71.124022,
  title = {Stability of phantom wormholes},
  author = {Lobo, Francisco S. N.},
  journal = {Phys. Rev. D},
  volume = {71},
  issue = {12},
  pages = {124022},
  numpages = {9},
  year = {2005},
  month = {Jun},
  publisher = {American Physical Society},
  doi = {10.1103/PhysRevD.71.124022},
  url = {https://link.aps.org/doi/10.1103/PhysRevD.71.124022}
}

@article{PhysRevD.87.067504,
  title = {Modified-gravity wormholes without exotic matter},
  author = {Harko, Tiberiu and Lobo, Francisco S. N. and Mak, M. K. and Sushkov, Sergey V.},
  journal = {Phys. Rev. D},
  volume = {87},
  issue = {6},
  pages = {067504},
  numpages = {5},
  year = {2013},
  month = {Mar},
  publisher = {American Physical Society},
  doi = {10.1103/PhysRevD.87.067504},
  url = {https://link.aps.org/doi/10.1103/PhysRevD.87.067504}
}

@article{Poisson:1995sv,
    author = "Poisson, Eric and Visser, Matt",
    title = "{Thin shell wormholes: Linearization stability}",
    eprint = "gr-qc/9506083",
    archivePrefix = "arXiv",
    doi = "10.1103/PhysRevD.52.7318",
    journal = "Phys. Rev. D",
    volume = "52",
    pages = "7318--7321",
    year = "1995"
}

@article{Gonzalez:2008wd,
    author = "Gonzalez, J. A. and Guzman, F. S. and Sarbach, O.",
    title = "{Instability of wormholes supported by a ghost scalar field. I. Linear stability analysis}",
    eprint = "0806.0608",
    archivePrefix = "arXiv",
    primaryClass = "gr-qc",
    reportNumber = "UMSNH-IFM-F-2008-18",
    doi = "10.1088/0264-9381/26/1/015010",
    journal = "Class. Quant. Grav.",
    volume = "26",
    pages = "015010",
    year = "2009"
}

@article{Pastén_2026,
doi = {10.1088/1361-6382/ae73df},
url = {https://doi.org/10.1088/1361-6382/ae73df},
year = {2026},
month = {jun},
publisher = {IOP Publishing},
volume = {43},
number = {11},
pages = {115014},
author = {Pastén, Erick and Bosquez, Marco and Cruz, Norman},
title = {Null Raychaudhuri equation and the impossibility of traversable wormholes in unimodular gravity},
journal = {Classical and Quantum Gravity}
}

@misc{alencar2026,
      title={The scalar--Maxwell--$\Lambda(x)$ system: Wormhole spacetimes without nonlinear electrodynamics in unimodular gravity}, 
      author={G. Alencar and T. M. Crispim},
      year={2026},
      eprint={2603.30003},
      archivePrefix={arXiv},
      primaryClass={gr-qc},
      url={https://arxiv.org/abs/2603.30003}, 
}

@misc{aguilarpérez2026,
      title={Spatial curvature in Unimodular Gravity}, 
      author={Gilberto Aguilar-Pérez and Miguel Cruz and Samuel Lepe},
      year={2026},
      eprint={2605.16751},
      archivePrefix={arXiv},
      primaryClass={gr-qc},
      url={https://arxiv.org/abs/2605.16751}, 
}

@misc{Alencar:2026ffr,
    author = "Alencar, G. and Borralho, V. H. U.",
    title = "{Dymnikova Black Holes in Unimodular Gravity: Maxwell Sources and Vacuum Contributions}",
    eprint = "2605.15255",
    archivePrefix = "arXiv",
    primaryClass = "gr-qc",
    month = "may",
    year = "2026"
}

@article{Alvarenga:2025nwe,
  author = {Alvarenga, Marcelo H. and Fabris, J{\'u}lio C.},
  title = {{Anisotropic Bianchi-I cosmological model in non-conservative unimodular gravity}},
  eprint = {2511.20486},
  archiveprefix = {arXiv},
  primaryclass = {gr-qc},
  doi = {10.1088/1361-6382/ae4d9e},
  journal = "Class. Quant. Grav.",
  volume = {43},
  pages = {055013},
  year = {2026},
}

@article{Godel:2025,
author = {Raimarda, R. and Santos, A. F. and Bufalo, R.},
title = {G\"{o}del-type universes in unimodular gravity},
journal = {International Journal of Modern Physics D},
volume = {34},
number = {12},
year = {2025},
doi = {10.1142/S0218271825500488},
URL = {https://doi.org/10.1142/S0218271825500488},
eprint = {https://doi.org/10.1142/S0218271825500488}
}
 
\end{document}